\documentclass[useAMS,usenatbib   
]{mn2e}   
\usepackage{times}   
\usepackage{graphicx}   
\usepackage{amsmath}   
\usepackage{natbib}   
\usepackage{color,ulem}  
\bibpunct{(}{)}{;}{a}{}{,}

\title[Ultra-Compact Dwarfs around NGC\,3258 in Antlia]{Ultra-Compact Dwarfs around    
NGC\,3258 in the Antlia cluster \thanks{Based on observations collected at the Cerro  
Tololo Interamerican Observatory (CTIO); observations obtained at the Gemini   
Observatory, which is operated by the Association of Universities for Research in   
Astronomy, Inc., under a cooperative agreement with the NSF on behalf of the Gemini   
partnership: the National Science Foundation (United States), the Science and   
Technology Facilities Council (United Kingdom), the National Research Council (Canada),   
CONICYT (Chile), the Australian Research Council (Australia), Minist\'erio da Ciencia e   
Tecnologia (Brazil) and Ministerio de Ciencia, Tecnolog\'ia e Innovaci\'on Productiva   
(Argentina); and observations carried out at the European Southern Observatory, Paranal   
(Chile), programme 71.B-0122(A).}}  
  
\author[Juan Pablo Caso et al.]    
{Juan Pablo Caso$^{~1,2}$\thanks{E-mails:\,jpceda@carina.fcaglp.unlp.edu.ar\,(JPC);   
\,lbassino@fcaglp.unlp.edu.ar\,(LPB);\,tom@astro-udec.cl\,(TR);   
\,asmith@fcaglp.fcaglp.unlp.edu.ar\,(ASC);\,favio@fcaglp.fcaglp.unlp.edu.ar\,(FF)}, Lilia P. Bassino$^{~1,2}$, Tom Richtler$^{~3}$, 
\newauthor
Anal\'ia V. Smith Castelli$^{~1,2}$ and Favio Faifer$^{~1,2}$\\   
$^{1}$Facultad de Ciencias Astron\'omicas y Geof\'isicas de la Universidad Nacional de La Plata,    
and \\ Instituto de Astrof\'isica de La Plata (CCT La Plata -- CONICET, UNLP), Paseo del Bosque S/N,  
B1900FWA La Plata, Argentina\\   
$^{2}$Consejo Nacional de Investigaciones Cient\'ificas y T\'ecnicas, Rivadavia 1917, C1033AAJ  
Ciudad Aut\'onoma de Buenos Aires, Argentina\\   
$^{3}$Departamento de Astronom\'ia, Universidad de Concepci\'on, Casilla 160--C, Concepci\'on, Chile}  
\begin{document}   
   
\date{Accepted . Received ; in original form }   
   
\pagerange{\pageref{firstpage}--\pageref{lastpage}} \pubyear{2011}   
   
\maketitle   
   
\label{firstpage}   
   
\begin{abstract}  
We present the first compact stellar systems with luminosities in the range of 
ultra-compact dwarfs (UCDs), discovered in the Antlia galaxy cluster ($-10.5 < M_V < -11.6$). 
The magnitude limit between UCDs and globular clusters (CGs) is discussed.   
By means of imaging from VLT (FORS1), CTIO (MOSAIC), and the HST (ACS) archive,   
eleven UCDs/bright GCs are selected on the basis of photometry   
and confirmed as Antlia members through radial velocities measured on new GEMINI   
(GMOS-S) spectra. In addition, nine UCD candidates are identified taking into account   
properties derived from their surface brightness profiles. All of them, members and   
candidates, are located in the proximity of NGC\,3258, one of the two brightest elliptical    
galaxies in the cluster core. Antlia UCDs in this sample present absolute    
magnitudes fainter than $M_V \sim -11.6\,{\rm mag}$ and most of them have colours within   
the blue GC range, falling only two within the red GC range. Effective radii measured for   
the ones lying on the ACS field are in the range $R_{\rm eff} = 3 - 11$\,pc and   
are similar to equivalent objects in other clusters, obtained from the literature.   
The UCD sample shares the same behaviour on the size-luminosity plane: a linear relation   
between $R_{\rm eff}$ and $M_V$ is present for UCDs brighter than $M_V \sim -10.5$ -- $-11\,{\rm mag}$   
while no trend is detected for fainter ones, that have an approximately constant   
$R_{\rm eff}$. The projected spatial distribution of UCDs, GCs and X-ray emission   
points to an ongoing merger between two Antlia groups, dominated by NGC\,3258 and NGC\,3268.   
Nuclei of dwarf elliptical galaxies and blue UCDs share the same locus on the colour-magnitude   
diagram, supporting the hypothesis that some blue UCDs may be remnants of stripped 
nucleated dwarfs.   
   
\end{abstract}   
   
\begin{keywords}   
galaxies: star clusters -- galaxies: photometry -- galaxies: nuclei        
\end{keywords}
   
\section{Introduction}   

Ultra-Compact Dwarfs (UCDs) is the name that has been assigned (first by \citealt{dri00}) 
to an apparently  new class of compact objects with masses and luminosities ranging between 
globular clusters (GCs) and dwarf galaxies. Their origin and nature are not yet completely 
understood. 
Most authors assume that they in fact may have various formation channels 
\citep[e.g.][]{hil09,nor11,chil11}. The first UCDs were discovered in the proximity of 
NGC\,1399, the dominant galaxy of the Fornax cluster \citep{min98,hil99}. Afterwards, UCDs 
have also been found in other galaxy clusters/groups \citep{mie07,evs08,gre09,mad10,mis11,
rej07,hau09,dar11,mad11}. Although there is no generally accepted definition of UCD 
luminosities, \citet{hil09} suggests a $V$ absolute magnitude range of $-13.5 < M_V < -11$.

The defining criterion of what is a UCD varies for different authors. Metallicity, radius,
luminosity or mass-to-light ratio ($M/L$) tresholds have been proposed, according to the behaviour 
of these properties in compact objects \citep[e.g.][]{mie06,mie08,nor11}. The presence of multiple 
stellar populations, as found in $\omega$\,Cent \citep[e.g.][]{and09} is other possibility. 
\citet{bro11} employ the effective radius ($R_{\rm eff}$) as the property which separates UCDs from 
GCs. Setting a limit of $R_{eff}$ =10\,pc, they include objects as faint as $M_V = -9\,{\rm mag}$.
\citet{mie08} and \citet{dab08} suggested $2\,\times10^6\,M_{\odot}$ as a limiting mass.

Although some objects were found in merger remnants of intermediate age like W3 in 
NGC\,7252 \citep{mar04}, most UCDs present an old stellar population (t $\sim10$\,Gyr, 
\citealt{mie06,evs07}) and colours within a similar range as GCs.  Their $R_{\rm eff}$ 
can reach up to $\sim100$\,pc, and their brightness profiles present 
nuclear and halo components. However, UCDs usually present $7 < R_{\rm eff}< 30 $\,pc 
\citep{mie07,mie08,evs08,chi11}. Dynamical masses range between 
$2\times10^6 <$ M $<10^8M_{\odot}$ \citep{mie08,chil11}. Some studies find that mass-to-light
ratios ($M/L$) can assume values twice as high as those of  Galactic GCs of similar 
metallicity \citep{evs07,mie08,tay10}. 
There is  not a general agreement about the presence of dark matter haloes in these objects 
\citep[e.g.][]{chil11}. \citet{fra11} analysed the internal kinematics of UCD3 in Fornax with 
spatially resolved spectroscopy, and they did not find evidence for a dark matter component in 
this object. 

\begin{figure}   
\includegraphics[width=80mm]{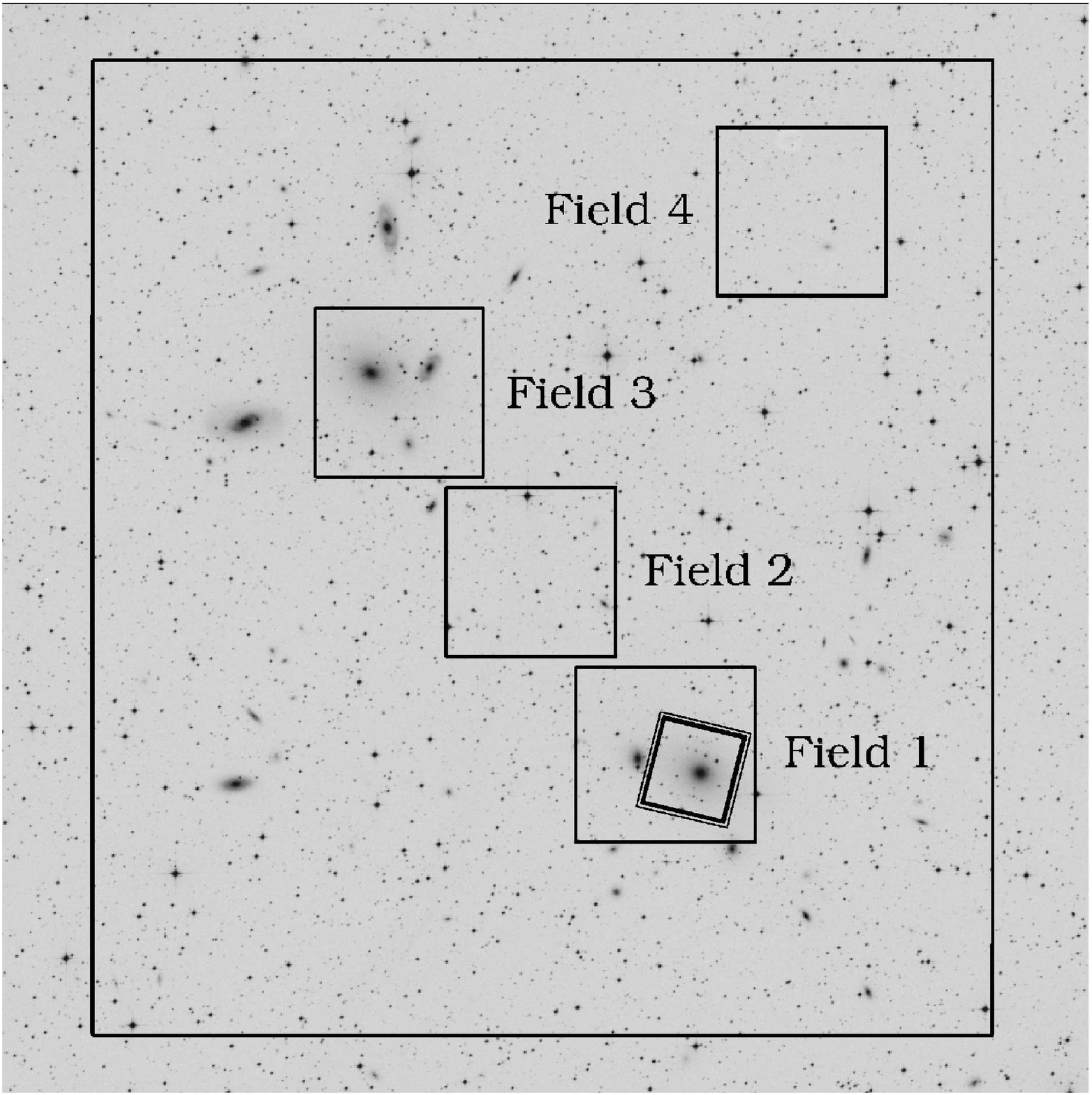}   
\caption{The four FORS1--VLT fields are overlaid on a DSS 
image of the central part of the Antlia cluster. The outer box indicates the MOSAIC--CTIO field 
(about $36\times36\,{\rm arcmin^2}$)}. The framed box indicates the
ACS field. {\bf At the adopted Antlia distance, $1\,arcsec$
corresponds to $\approx 170\,{\rm pc}$}. North is up, East to the left.   
\label{dss}   
\end{figure}

\subsection{Possible formation scenarios for UCDs}   

A considerable number of theories attempt to explain the existence of this    
 class of stellar systems. The most discussed ones are:   
\begin {enumerate}   
   
\item UCDs can be remnants of galaxies, for instance nucleated dwarf    
elliptical galaxies (dE,N), that had been disrupted by the tidal forces    
of massive galaxies (e.g. \citealt{bas94,bek01,goe08}). 

\item UCDs may be the result of the fusion of several young star clusters    
(e.g., \citealt{fel02,fel05}). 

\item UCDs can be the brightest members of the GC system (GCS)    
associated to a host galaxy (e.g., \citealt{hil09,nor11}).

\end {enumerate}   
More detailed information about formation scenarios has been given by    
\citet{hil09} and, more recently, by \citet{mis11} and \citet{nor11}.   
  
\subsection{The Antlia cluster}   

This work is part of the Antlia Cluster Project, that is devoted to study the    
different stellar systems of this cluster, from GCs \citep{dir03b,bas08} to the     
galaxy population \citep{smi08a,smi08b,smi12}.

The Antlia galaxy cluster, located in the Southern sky at low Galactic    
latitude ($l\sim\,19\degr$), is after Virgo and Fornax the nearest populous   
galaxy cluster. The central part of the cluster consists of two subgroups, 
each one dominated by a giant elliptical galaxy (NGC\,3258 and NGC\,3268) of 
comparable luminosity. Galaxies located in both subgroups present an elongated 
projected distribution, in the direction that joins the two giant ones. 

\citet{haw11} recently presented a study of the inner 12\,arcmin of the Antlia cluster, 
using XMM--Newton. They consider Antlia as `the nearest example of a galaxy cluster in 
an intermediate merger stage without a cool core'. While the early- to late-type galaxy 
ratio indicates an evolved system, the existence of two subgroups, which may also be 
present in the overall mass distribution, means that the total system has not yet 
completed its evolution. We might be witnessing the merging of two, rather evolved, compact 
groups/clusters.
\medskip    
   
Due to the influence that environmental conditions may have on the origin and dynamical 
evolution of the UCDs, the Antlia cluster is a very interesting system to study. 
Moreover, increasing the sample of analysed UCDs will help to understand their nature.

We present here the first results of the search for UCDs in the Antlia cluster, focusing
 on the surroundings of NGC\,3258. In this investigation, first, we analyse a preliminary 
sample that also includes the `supposedly' brightest GCs and will generally refer to both, 
bright GCs and UCDs, as `compact objects'.   
Then, we perform a more refined selection (see Sect. 4) of a specific `UCD sample' (Antlia 
members and candidates) and use it for the second part of this research.   
Preliminary results on the search for UCD candidates in Antlia have been given by   
\citet{cas09,cas10}.

This paper is structured as follows. Section\,2 describes the observations,   
reductions, and adopted criteria for the compact objects selection. In Section\,3   
we present and discuss the results regarding their colour-magnitude relation (CMR), 
size, size-luminosity relation, and we also compare magnitudes and colours of the   
compact objects with those obtained for a sample of Antlia dE,N nuclei.   
The final selection of UCD members and candidates is performed in Section\,4,   
that deals with their projected spatial distribution, colours, metallicities,   
as well as the Antlia global colour-magnitude diagram (CMD). Finally, a summary   
and the conclusions are provided in Section\,5.        
   
\section[]{Observations and reductions}   

\begin{figure}   
\includegraphics[width=84mm]{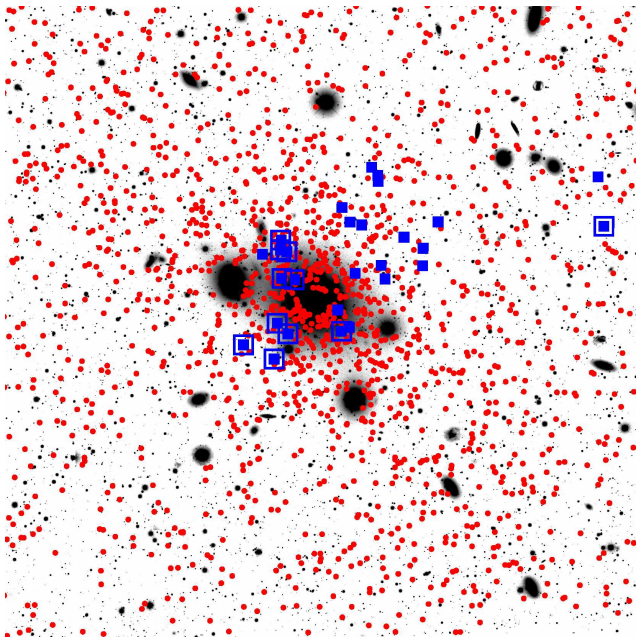}   
\caption{Projected spatial distribution for GC candidates with $T_1<23.6$ around NGC\,3258 (red filled
circles), and the spectroscopically observed sources (blue filled squares). The confirmed Antlia members, named with the acronym ACO,
are indicated with framed filled squares. The scale of the image is $\approx 20 \times 20 {\rm arcmin^2}$.
North is up, East to the left}   
\label{ran}   
\end{figure}

In this Section we describe the photometric and spectroscopic data, how the   
surface brightness profiles of Antlia dE galaxies have been obtained, and the   
identification and selection of the compact objects, i.e. UCDs and bright GCs.  
   
\subsection{Observational data}

\begin{figure}   
\includegraphics[width=84mm]{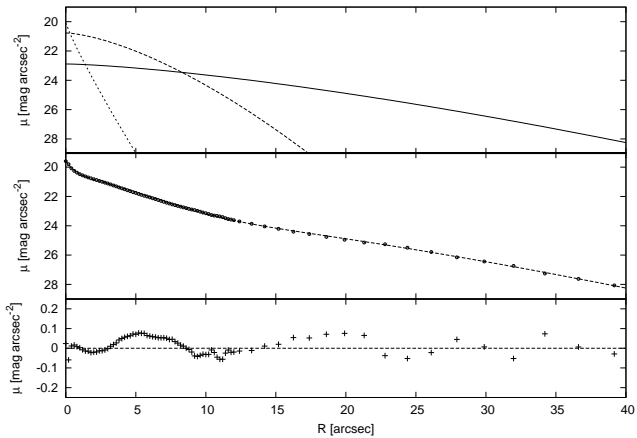}   
\caption{Three--component S\'ersic model fitted to the surface brightness profile   
of the galaxy FS90-177 ($V$ filter). In the upper panel are plotted individually the three 
S\'ersic components. The surface brightness profile obtained from ELLIPSE is superimposed to 
the sum of the three components in the medium panel, achieving a good agreement. The lower 
panel shows the residuals of the global fit. This is the only galaxy from our sample for 
which a two--component S\'ersic profile was needed to fit accurately the surface 
brightness profile.}   
\label{fs177}   
\end{figure}

The photometric observations used in this paper are obtained from three different   
sources.
Part of the material consists of FORS1--VLT images in the $V$ and $I$ bands (programme 
71.B-0122(A), PI B. Dirsch). These images correspond to four fields, two of them are centred 
on each one of the dominant galaxies, NGC\,3258 (Field\,1) and NGC\,3268 (Field\,3), the 
third one is located in the region between them (Field\,2), and the last one is located to 
the north-west direction (Field\,4, see Fig.\,\ref{dss}). 
  
We also use wide--field images that were taken with the MOSAIC camera mounted at the CTIO 4-m 
Blanco telescope (indicated by the outer box in Fig.\,\ref{dss}) during 4/5 April 2002. The 
Kron--Cousins $R$ and Washington $C$ filters were used, although the genuine Washington system 
uses $T_1$ instead of $R$. However, \citet{gei96} showed that the Kron--Cousins $R$ filter is 
more efficient than $T_1$ and that there is only a small colour term and zero-point difference 
between magnitudes in both filters (we use $R - T_1 = 0.02$ from \citealt{dir03b}). These images 
were originally used to perform the first study of the GCSs of NGC\,3258 and NGC\,3268 in Antlia 
\citep{dir03b}. 

In addition, an ACS field on NGC\,3258 (indicated by the framed box in 
Fig.\,\ref{dss}) observed with the $F814$\,filter was obtained from the 
Hubble Space Telescope Data Archive. This image is the composite of four 
570\,s exposures and corresponds to the programme 9427 (PI: W. E. Harris). 

We have also obtained GEMINI--GMOS multi-object spectra for compact objects located    
in  five Antlia fields (programmes GS-2008A-Q-56, PI T. Richtler; GS-2009A-Q-25, PI L. P. 
Bassino). The masks designed for these programmes were devoted to the study of the general 
population of the cluster, not only the compact objects. In all cases, the B600\_G5303 grating blazed 
at $5000\,\mathrm{\AA}$ was used, with three different central wavelengths (5000, 5050, and 
$5100\,\mathrm{\AA}$) in order to fill in the CCD gaps. A slit width of 1\,arcsec was selected. 
Considering an average seeing of 0.5 -- 0.6\,arcsec, this configuration gives a wavelength    
coverage of $3300 - 7200\,\mathrm{\AA}$ depending on the positions of the slits, and a 
spectral resolution of $\sim 4.6\,\mathrm{\AA}$. The total exposure times ranged between 2 
and 3.3\,h. Individual calibration flats and CuAr arc spectra were obtained for each 
exposure during the programme time, in order to avoid small variations that could be 
introduced by flexion of the telescope.
Figure\,\ref{ran} shows the projected spatial distribution of GC candidates with 
$T_1<23.6$ around NGC\,3258 (circles), and the spectroscopically observed sources 
(squares).     

\subsection{Photometry and source selection}   
    
In order to work in an homogeneous way and not to lose any UCD candidate that 
might has been discarded in the past as too bright for being a GC, we redo the 
basic photometry for the FORS1 and MOSAIC data.  
    
The FORS1 photometry was performed with DAOPHOT within IRAF, using a spatially 
variable point-spread function (PSF). The point source selection was based on 
the $\chi$ and sharpness parameters from the ALLSTAR task. We refer to \citet{bas08} 
for more details on the observations and the calibration equations applied to 
obtain colours and magnitudes in the standard system.    

\begin{figure*}   
\includegraphics[width=53mm]{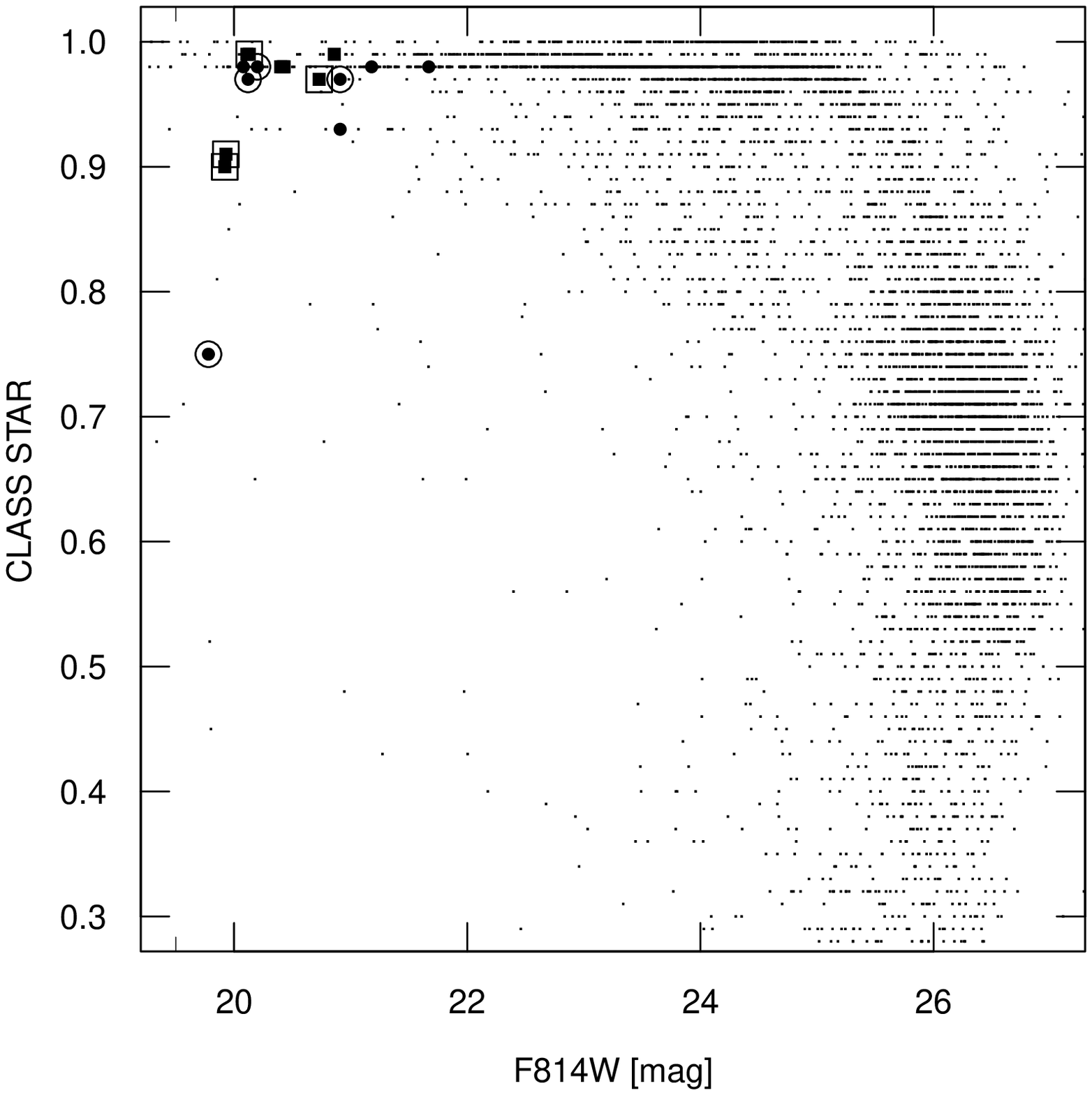}   
\includegraphics[width=53mm]{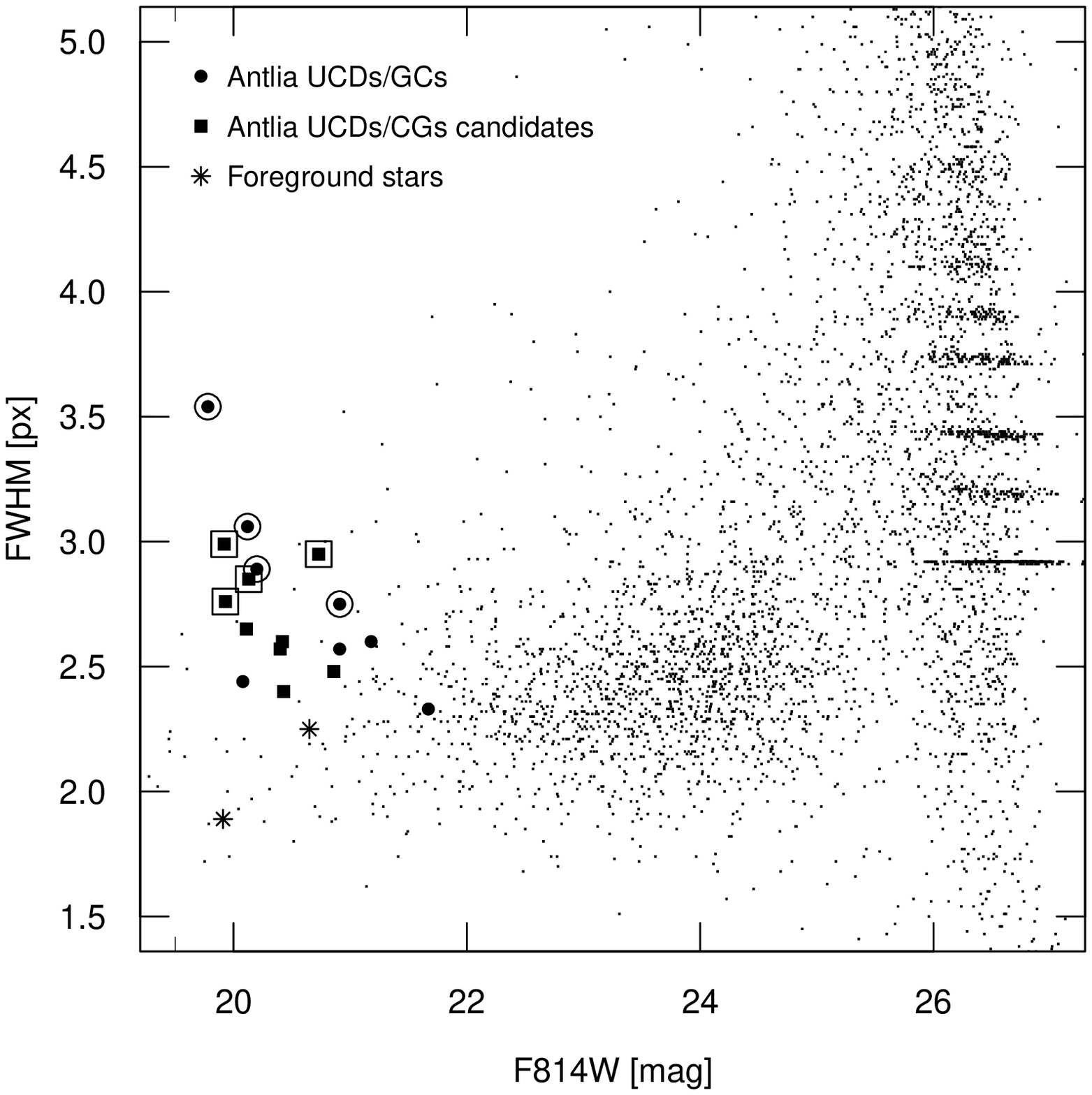}
\includegraphics[width=53mm]{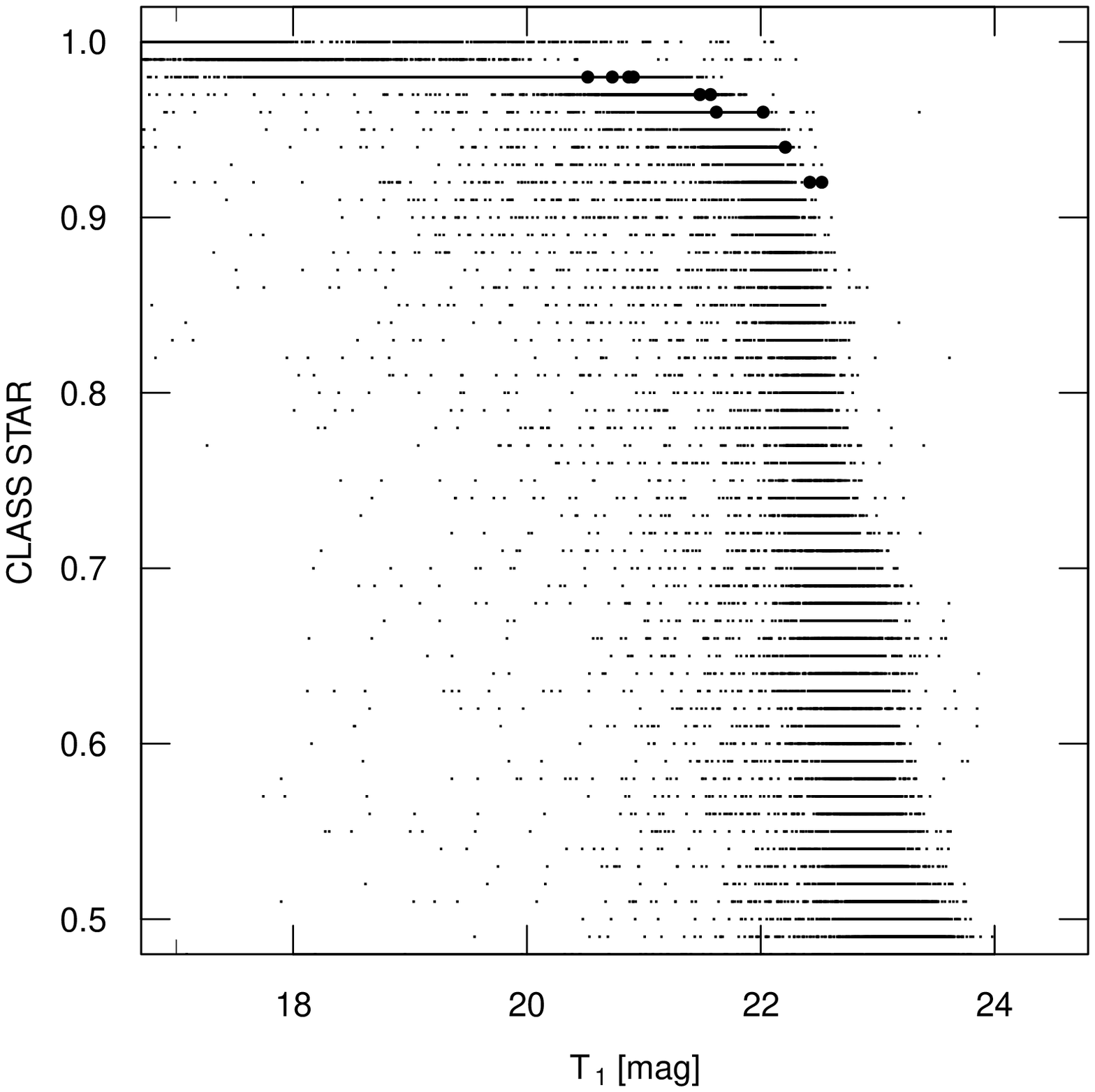}   
\caption{{\bf Left-hand panel}: SExtractor's CLASS$\_$STAR as a function of    
the instrumental $F814$\,magnitude for all the compact objects in the NGC\,3258   
ACS field.    
Filled circles represent confirmed Antlia members. Filled squares indicate    
marginally resolved UCD candidates (see Section\,3.3.). Framed symbols distinguish those     
with $R_{\rm eff}>5\,pc$ among confirmed members and candidates.    
{\bf Middle panel}: SExtractor's FWHM as a function of the instrumental   
$F814$\,magnitude for compact objects with CLASS$\_$STAR$>$0.4 in the same field.     
Symbols are the same as in the left-hand panel. Asterisks indicate two    
confirmed foreground stars.{\bf Right-hand panel}: SExtractor's CLASS$\_$STAR as a function of    
the instrumental $T_1$\,magnitude for objects in the MOSAIC field. Filled circles represent 
confirmed Antlia members. }   
\label{sex}   
\end{figure*}

The MOSAIC photometry was also performed with DAOPHOT. The extended galaxy light 
was subtracted, using a median ring filter with an inner radius of 9 arcsec and an 
outer radius of 11 arcsec. This facilitates point source detection, without 
modifying the results of the subsequent photometry \citep{dir03a,dir03b,bas06a,bas06b}.  
The software SExtractor \citep{ber96} was applied to the sky-subtracted $R$ image 
to obtain an initial selection of point sources. The software was set so as to 
consider a positive detection every group of, at least, five connected pixels 
above a threshold of $1.5\sigma$ (DETECT\_MINAREA and DETECT\_TRESH parameters, 
respectively). At the Antlia distance, that we adopt as approximately $35$\,Mpc 
(distance modulus $(m-M) = 32.73$, \citealt{dir03b}), even the larger UCDs are seen 
as point sources on our MOSAIC images. Thus, to select point sources we use the 
star/galaxy classifier from SExtractor, through the CLASS$\_$STAR parameter, 
that takes values close to one for point sources and close to zero for extended 
sources (see the right-hand panel in Fig.\,\ref{sex}). All the objects with 
CLASS$\_$STAR$<0.6$ are rejected. The aperture 
photometry was performed using the task PHOT. Afterwards, a spatially variable 
PSF was built, employing about a hundred bright stars, well distributed over the 
whole field. The final point source selection was based on the $\chi$ and sharpness 
parameters from the ALLSTAR task, and the aperture corrections and calibration 
equations were obtained from \citet{dir03b}.  

In the reduction of the ACS data, the surface brightness profile of the galaxy was 
obtained with the task ELLIPSE and the corresponding synthetic galaxy generated with 
BMODEL (both IRAF tasks) was subtracted from the original image. A first source 
selection was made with SExtractor, considering a positive identification every 
detection of at least three connected pixels above a threshold of $1.5\sigma$.
In this case, as UCDs may be marginally resolved on ACS images at the Antlia distance,   
we decided to reject all  objects with CLASS$\_$STAR$<0.4$ (see the left-hand panel in 
Fig.\,\ref{sex}). Again, aperture photometry 
was performed  using the task PHOT, and the PSF was built with the DAOPHOT/IRAF homonymous 
task, selecting bright stars well distributed on the field. In the following, we 
will keep just the instrumental $F814$ magnitudes.
A list of UCD and bright GC candidates was then compiled with the objects that 
satisfy the source selection criteria in the three photometric data sets (from FORS1, 
MOSAIC, and ACS images), and have magnitudes in the range $-13.5 < M_V < -10.5$. At 
the adopted Antlia distance, it corresponds to an apparent magnitude range 
$19.2 < V < 22.2$. Considering the mean colour for early-type galaxies $V-R\sim0.6$ 
given by \citet{fuk95} also valid for UCDs, the previous magnitude range corresponds 
to $18.6 < T_1 < 21.6$ in the Washington photometric system. This list should be 
considered as a preliminary selection, namely the compact objects, as we will refine 
it later when sizes are also taken into account to separate the UCDs.

\subsection{Surface brightness profiles of dwarf elliptical galaxies}
On the basis of the hypothesis that links bright compact objects (i.e. UCDs) with the 
nuclei of dE,N galaxies, it was tried to isolate the nuclei of dE,N in order to compare 
their photometric properties with those of UCDs.
A list of Antlia dEs was compiled considering the members confirmed by radial velocities 
 \citep{smi08a,smi12} as well as galaxies labelled as `definite members' by \citet{fer90} 
(type-1, in their classification). Even though \citet{fer90} only used photometric galaxy 
properties as membership criteria of the galaxies, subsequent spectroscopic observations 
indicated that the vast majority of type-1 galaxies are in fact Antlia members 
\citep[][and references  therein]{smi12}. Ten dwarf galaxies from this member list are 
located on our FORS1--VLT images, and for nine of them we could obtain luminosity profiles.   
The remaining one (FS90-109, where FS90 corresponds to the identification given by 
\citealt{fer90}) is very close to a saturated star, making it difficult to obtain an 
acceptable fit. 

The following process was applied to get such galaxy profiles. First, the 
FORS1--VLT images were visually inspected to determine if the extended light of bright 
galaxies, present in the field, could affect the dwarf photometric measurements. In 
Field\,1 the luminosity profile of NGC\,3258 was obtained using the task ELLIPSE within 
IRAF, after masking all the bright objects in the field. Then, a model galaxy generated 
with the task BMODEL was subtracted from the original image. This second image, free of 
the extended light from NGC\,3258, was used to fit the luminosity profile of NGC\,3260, 
another bright galaxy present in this field. Once again, a model galaxy was generated with 
BMODEL and subtracted from the original image. The resultant image was used to obtain the 
surface brightness profile of NGC\,3258, repeating the iterative process three times for each 
filter. The final models of both bright galaxies were then subtracted from the original   
image. 

A similar process was implemented for Field\,3, subtracting the light from galaxies    
NGC\,3267, NGC\,3268 and FS90-175.    
The Fields\,2 and 4 do not contain bright galaxies. The correspoding background levels
were calculated for each image using its mode. This was done applying a rejection level 
of three times the dispersion to eliminate outliers, and the mode was re-obtained. 
This iterative process was repeated 50 times.  
Also, for each dE the surrounding background was searched for possible gradients.   
A plane was fitted when necessary, in order to take into account possible variations of 
the background in the region where dwarfs were located, originated in the subtracted bright 
galaxies luminosity profiles, or the contribution of bright saturated stars.   

For every galaxy, the ELLIPSE task was used to obtain the luminosity profiles in both $V$ 
and $I$ filters. These dE profiles are usually characterized by S\'ersic models \citep{ser68},    
\citep[e.g.][]{geh02,buz12}. Considering surface brightness units (mag\,arcsec$^{-2}$), an 
alternative form for the equation of such models is:

\begin{equation}  
\mu(r) = \mu_0 + 1.0857\,\left(\frac{r}{r_0}\right)^N,  
\end{equation}  
  
\noindent where $\mu_0$ is the central surface brightness,   
 $r_0$ is a scale parameter and $N$ is the S\'ersic index.   
If this is compared with the original form of the S\'ersic law,   
  
\begin{equation}  
I(r) = I_e\,exp\,\left\{b_n\left[\left(\frac{r}{r_e}\right)^{\frac{1}{n}}-1\right]\right\}  
\end{equation}  
  
\noindent it follows that N$=1/$n, and\\[-8mm]  
  
\begin{equation}  
\mu_0 = \mu_e-1.0857b_n\\[-8mm]  
\end{equation}  
  
\begin{equation}  
r_0 = b_n^{-n}r_e  
\end{equation}  
  
For each dwarf galaxy, its calibrated surface brightness profile was fitted by a S\'ersic law. 
The residuals between the observed and fitted profiles were obtained in all cases. For those 
galaxies where one S\'ersic law did not provided an acceptable fit, due to the presence of a 
nucleus, a two--component fit was performed. In these cases, first a single S\'ersic profile 
was fitted to the outer region of the galaxy, providing a very good fit to the halo component. 
Then, it was subtracted from the intensity profile of the galaxy, resulting in a profile 
intrinsically dominated by the nuclear component. Nuclei present  half-light radii of only a 
few parsecs (e.g., \citealt{cot06}), so they appear as point sources on our FORS1 images.
 For this reason, the total flux of the nucleus was directly obtained from the luminosity profile, 
once the outer component had been subtracted. Then, the integrated magnitude of the nucleus was 
calculated from this flux.   

From this sample, only for one galaxy a single S\'ersic profile could reproduce its surface 
brightness profiles. The remaining eight had  nuclei, and seven were well described by    
compositions of a S\'ersic halo and a point-source-like nucleus. \citet{fer90} 
classified three of them as dE,N (FS90-87, FS90-162 and FS90-176) and three as dE   
(FS90-103, FS90-186 and FS90-196). For one, the existence of a nucleus was under 
debate (FS90-159). FS90-136 was the only one well fitted by a single S\'ersic profile, despite 
that \citet{fer90} classify it as a dE,N.  
The remaining dwarf is FS90-177. It is described by \citet{fer90} as d:E,N with total B magnitude 
$B_T=17.0$.
 In this case a two--component S\'ersic profile was needed to fit accurately the surface 
brightness profile of the galaxy extended component. The process was similar to that followed in the 
previous cases. Every component was fitted independently to the profile, after subtracting the profiles 
obtained for the outer ones. The implementation of an additional component must be carefully analysed, in 
the sense that it could have no physical support and be described just an image defect. However, as can 
be seen in Fig.\,\ref{fs177} where the results of the fits for this galaxy are shown, this additional 
component was needed to obtain an acceptable fit.  

\subsection{Spectroscopic data}   

Data reduction was performed using the GEMINI.GMOS package within IRAF. A master 
bias was constructed for each observing run using approximately 20 
bias images, obtained from the Gemini Science Archive (GSA).

 These images are selected as the ones   
taken as close as possible to the science ones. In order to remove cosmic rays   
from science images, the FWHM of a Gaussian fit to the lines in the CuAr arc spectra   
was measured. Then, this value was used as a minimum rejection limit in the GSREDUCE task.   
The wavelength calibration was obtained with the task GSWAVELENGTH. The list of wavelength   
calibration lines was constructed from the default list provided by the GMOS package, after   
removing the lines that in a visual inspection of our CuAr arc images appeared to be faint,   
extremely close to another line, or with an unusual profile. Then, images were rectified   
and wavelength-calibrated with GSTRANSFORM task. For all the spectra, individual exposures   
were combined to achieve a higher S/N. The trace of every object was obtained from the  
combined images, and then used to extract spectra from the individual exposures. The 
y-positions in the slit for the individual exposures were the same for each object.   

Radial heliocentric velocities for the compact objects in the GMOS fields    
were measured using the task FXCOR in the NOAO.RV package within IRAF. For this purpose,   
templates were obtained from the single stellar population (SSP) model spectra at the MILES    
website (http://www.iac.es/proyecto/miles, \citealt{san06}), considering a SSP    
with metallicity [M/H]=\,-0.71, a unimodal initial mass function with a slope    
of $1.30$, and ages of 8 and 10\,Gyr. The wavelength coverage of the templates    
is $4200 - 7300\,\mathrm{\AA}$, and the spectral resolution is    
$2.3\,\mathrm{\AA}$ FWHM. In all cases the 10\,Gyr template provided a slightly   
better correlation.  

\begin{table*}   
\begin{minipage}{110mm}   
\begin{center}   
\caption{Basic properties of confirmed Antlia compact objects close to NGC\,3258. Visual absolute
magnitudes were calculated considering a distance modulus of $(m-M) = 32.73$ \citep{dir03b}}    
\label{table1}   
\begin{tabular}{@{}cc@{}c@{}c@{}c@{}cc@{}c@{}c@{}c@{}ccccccr@{}c@{}l@{}}   
\hline   
\\   
\multicolumn{1}{c}{ID}&\multicolumn{5}{c}{RA(J2000)}&\multicolumn{5}{c}{DEC(J2000)}&\multicolumn{1}{c}{V$_0$}&  
\multicolumn{1}{c}{(V\,-\,I)$_0$}&\multicolumn{1}{c}{$M_V$}&\multicolumn{1}{c}{(T$_1$)$_0$}
&\multicolumn{1}{c}{(C\,-\,T$_1$)$_0$}&\multicolumn{3}{c}{RV$_{\rm hel}$}\\   
\multicolumn{1}{c}{}&\multicolumn{5}{c}{hh mm ss}&\multicolumn{5}{c}{dd mm ss}&\multicolumn{1}{c}{mag}&  
\multicolumn{1}{c}{mag}&\multicolumn{1}{c}{mag}&\multicolumn{1}{c}{mag}&\multicolumn{1}{c}{mag}
&\multicolumn{3}{c}{km\,s$^{-1}$}\\   
   
\hline   
\\   
ACO\,1&$10$&$\,$&$28$&$\,$&$08.1$&$-35$&$\,$&$33$&$\,$&$57$&$-$&$-$&$-$&$22.12$&$1.48$&$3037$&$\pm$&$148$\\   
ACO\,2&$10$&$\,$&$28$&$\,$&$49.2$&$-35$&$\,$&$37$&$\,$&$18$&$21.66$&$1.13$&$-11.07$&$20.98$&$1.83$&$2811$&$\pm$&$67$\\  
ACO\,3&$10$&$\,$&$28$&$\,$&$56.6$&$-35$&$\,$&$35$&$\,$&$39$&$22.31$&$0.99$&$-10.42$&$21.65$&$1.69$&$2577$&$\pm$&$66$\\ 
ACO\,4&$10$&$\,$&$28$&$\,$&$57.6$&$-35$&$\,$&$37$&$\,$&$23$&$22.52$&$0.99$&$-10.21$&$22.10$&$1.35$&$2800$&$\pm$&$174$\\  
ACO\,5&$10$&$\,$&$28$&$\,$&$57.8$&$-35$&$\,$&$34$&$\,$&$45$&$21.48$&$1.05$&$-11.25$&$20.90$&$1.57$&$2401$&$\pm$&$46$\\  
ACO\,6&$10$&$\,$&$28$&$\,$&$58.6$&$-35$&$\,$&$35$&$\,$&$37$&$21.17$&$0.98$&$-11.56$&$20.52$&$1.52$&$2725$&$\pm$&$85$\\  
ACO\,7&$10$&$\,$&$28$&$\,$&$58.8$&$-35$&$\,$&$34$&$\,$&$23$&$21.49$&$1.01$&$-11.24$&$20.86$&$1.56$&$3038$&$\pm$&$80$\\ 
ACO\,8&$10$&$\,$&$28$&$\,$&$58.9$&$-35$&$\,$&$34$&$\,$&$45$&$22.34$&$1.06$&$-10.39$&$21.55$&$1.88$&$2253$&$\pm$&$73$\\ 
ACO\,9&$10$&$\,$&$28$&$\,$&$59.3$&$-35$&$\,$&$37$&$\,$&$03$&$22.87$&$0.90$&$-9.86$&$22.31$&$1.37$&$1668$&$\pm$&$108$\\   
ACO\,10&$10$&$\,$&$28$&$\,$&$59.8$&$-35$&$\,$&$38$&$\,$&$11$&$22.83$&$0.93$&$-9.90$&$22.25$&$1.54$&$2616$&$\pm$&$166$\\ 
ACO\,11&$10$&$\,$&$29$&$\,$&$04.6$&$-35$&$\,$&$37$&$\,$&$44$&$22.21$&$1.02$&$-10.52$&$21.54$&$1.63$&$1997$&$\pm$&$131$\\   
\hline   
\end{tabular}    
\end{center}    
\end{minipage}   
\end{table*}   
   
\subsection{Selected sample of compact objects}   
   
Following our previous studies on the galaxy populations of the Antlia cluster    
\citep{smi08a,smi12}, objects with radial velocities in the range $1200 - 4200$\,km\,s$^{-1}$     
 will be considered as Antlia members. As result of our    
measurements, eleven compact objects are confirmed as cluster members and listed in Table\,1.     
As they have never been catalogued before, we use the acronym `ACO' for Antlia Compact Object,    
and give their J2000 coordinates, extinction corrected $V,I$,  Washington $C,T_1$ photometry,    
and heliocentric radial velocities. Extinction corrections are applied to magnitudes and colours   
in the rest of this paper, and are described in \citet{bas08} for the $V,I$ data and \citet{dir03b}   
for the Washington data.   
The sample listed in Table\,1 includes five compact objects with magnitudes slightly fainter than the   
adopted luminosity limit (see Section\,2.2). We prefer to keep them in the selected sample because they   
are identified as Antlia members by GMOS radial velocities. There is no available V,I photometry
for the object ACO\,1, because it is located outside the VLT fields.
Framed squares in Fig.\,\ref{ran} indicate the confirmed Antlia members.     
   
The left-hand panel of Fig.\,\ref{sex} shows the CLASS$\_$STAR parameter for the    
compact objects detected in the ACS field ( from ACO\,2 to ACO\,9) as a function of the    
instrumental $F814$\,magnitude estimated by SExtractor.   
These magnitude estimations are used    
here just for illustrative purposes and will not be considered in any further analysis.    
The plot presents the usual structure, with objects with CLASS$\_$STAR close    
to 1.0 covering the whole magnitude range, and doubtful point sources    
($0.4 <$ CLASS$\_$STAR $< 0.7$) mostly located in the faint regime.    
Filled circles represent the Antlia members confirmed by radial velocity measurements    
(see Table\,1).  
Their CLASS$\_$STAR values are higher than $0.90$ except one compact member   
with CLASS$\_$STAR$ \sim$\,0.75, but in all cases this parameter is larger than the lower limit   
CLASS$\_$STAR = 0.4 adopted for the selection. This gives us confidence that our selection criteria   
are not introducing any tendency in the sample.   
  
The middle panel exhibits the FWHM in pixel units as a function of the instrumental   
$F814$\,magnitude, both estimated by SExtractor, for the compact objects in the ACS field.    
Symbols are as in the left-hand panel. Filled circles represent Antlia members, and asterisks   
represent foreground stars confirmed with radial velocities. Confirmed compact objects present   
FWHM higher than the foreground stars and the rest of bright point sources. These figures show that    
both, SExtractor's FWHM and CLASS$\_$STAR parameters, can be used as auxiliary tools for selecting   
UCD/GC candidates \citep{chi11,mis11}.

\section{Results}   
  
The eleven compact objects presented so far in this paper (UCDs and bright GCs, Table\,1) have 
a weighted mean radial velocity  and standard deviation of $2528\pm102\,{\rm km\,s}^{-1}$ and 
$339\,{\rm km\,s}^{-1}$, respectively. \citet{smi08a} measured for NGC\,3258 a radial velocity 
of $2689 \pm 50\,{\rm km\,s}^{-1}$,  while NED\footnote{This research has made use of the NASA/IPAC 
Extragalactic Database (NED) which is operated by the Jet Propulsion Laboratory, California Institute 
of Technology, under contract with the National Aeronautics and Space Administration.} gives 
$2792\pm28\,{\rm km\,s}^{-1}$. If we take into account the more recent measurement of NGC\,3258 
velocity, the difference with the compact objects mean velocity is within the dispersion range. 
This points to a physical association of these compact objects with the host galaxy.     

As it has been suggested that UCDs may have quite different origins, and that the environment   
is certainly playing an important role \citep[e.g.][]{hil09,nor11}, we will compare the   
characteristics of the Antlia UCDs/bright GCs and dE,N nuclei with those belonging to other galaxy   
clusters. That is, as literally said by \citet{hil11}: {\it `A promising way to learn more about the   
nature of UCDs is to study their global properties in galaxy clusters and compare them to those   
of other dwarf galaxies and rich globular cluster systems around central cluster galaxies'}.  
   
\begin{figure}   
\includegraphics[width=84mm]{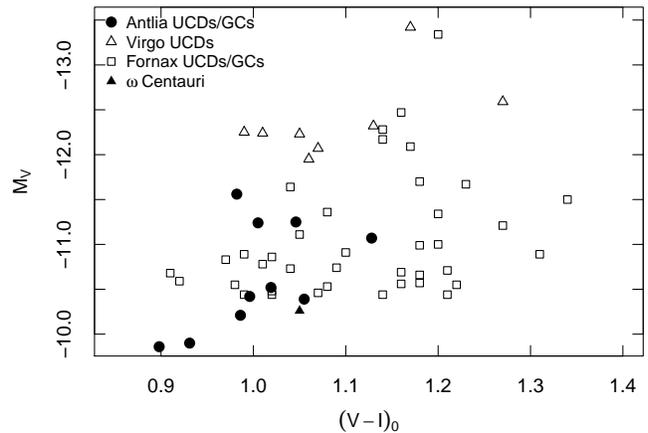}   
\caption{Colour-magnitude diagram of compact objects in Antlia (confirmed members), Fornax   
\citep{mie04}, and Virgo \citep{evs07}, and $\omega$ Centauri \citep[][2010 Edition]{har96}.}   
\label{cmd}   
\end{figure}

\subsection{Colour-magnitude diagram of compact objects}   

The CMD of the compact objects in the neighbourhood of    
NGC\,3258, confirmed as members of the Antlia cluster and with available $V,I$ photometry,    
is shown in Fig.\,\ref{cmd}. Only one object lacks photometric data (ACO\,1, see Table\,1)    
as it is located outside the VLT fields. We have also plotted compact objects found in    
nearby galaxy clusters (Virgo UCDs from \citealt{evs07}, Fornax UCDs/GCs from \citealt{mie04}) 
as well as $\omega$ Centauri \citep[][2010 Edition]{har96}, the only Galactic GC that presents 
an absolute $M_V$ magnitude similar to these extragalactic compact objects. More recently
published data from Fornax and Virgo compact objects, have not been included as they use different
photometric systems. We prefer to avoid photometric transformations that introduce more uncertaintes.
articles, have been published about Fornax and Virgo compact objects, but they used different
photometric systems, and we prefer to avoid the application of photometric transformations,
who introduce more uncertainties.  
   
None of the Antlia compact objects confirmed in this paper presents a luminosity     
as high as those of the brightest Virgo or Fornax UCDs. It is difficult to find   
a clear CMR due to the large dispersion, but a mild   
trend is present in the sense that brighter objects seem to be redder. We will   
come back later to the CMR once the Antlia dE,N nuclei have been added.   
  
All Antlia compact objects shown in Fig.\,\ref{cmd} are bluer than $V-I \sim 1.15$,   
while the Fornax sample reaches significantly redder colours. If we accept that   
compact objects brighter than $M_V \sim -11$    
are mostly UCDs, their mean $V-I$ colour is $1.04\pm0.03$ in Antlia,    
and $1.17\pm0.02$ in Fornax. In order to compare with colours of regular GCs (not just    
the brightest ones), we recall that for old GCs, the limit between metal-poor (`blue') and 
metal-rich (`red') ones is taken at $V-I \sim 1.05$ by \cite{bas08}, using the same VLT data set.     
Thus, only two out of the ten Antlia compact objects in the CMD are redder than    
this limit. Although the Virgo UCD sample is small, it can be seen that except the    
two very bright ones, the rest seem to be evenly distributed with respect to this      
blue/red GCs' colour limit. In this sense, \citet{bro11} analysed a larger sample of UCDs
around M\,87, and show that for magnitudes fainter than $M_i \approx -12.5$, these objects cover 
a narrow colour range, similar to the blue M\,87 GCs. Regarding the sample of Fornax compact 
objects, they cover the whole GC colour range, that is approximately $0.8 < (V-I) < 1.4$ 
\citep[e.g.][]{lar01,bas08}.    
   
However, this cannot be taken yet as an evidence that our Antlia compact objects are    
particularly blue. On one side, we are dealing with a small sample. On the other side,    
if we compare the GCSs associated to both dominant galaxies in Antlia, NGC\,3258 and    
NGC\,3268, the system around NGC\,3258 has a smaller fraction of red GCs and it does not    
extend to redder colours as much as the NGC\,3268 one does. Then, it may be just a    
consequence of the whole GC colour distribution being bluer too. In this case, one might    
wonder whether these UCDs around NGC\,3258 are closely related to an extension of the GCS     
towards brighter luminosities. Before sustaining such conjecture, in the next Section   
we will compare Antlia compact objects with nuclei of same cluster dE,N galaxies, and also   
come back to the Antlia CMD in Section\,4.4.  
     
\subsection{Luminosity profiles and nuclei of dE galaxies}  
As said above, one of the hypotheses for the origin of UCDs states that they could 
be remnants of dE,N galaxies, that had been disrupted by the tidal forces of massive 
galaxies. In this Section we describe the results of   
the study of nine dE galaxies located on the Antlia FORS1--VLT fields and their   
nuclei when present, with the aim of comparing the photometric properties of our
UCDs and dE nuclei.  
  
Table\,2 lists the dwarf galaxies analysed in the present work, the parameters of the S\'ersic   
models fitted to their luminosity profiles in both filters $V$ and $I$, and the total magnitude of   
each component.   
Besides the nucleus, FS90-177 presents an extended halo and an inner component,   
with similar central surface brightness but more concentrated towards the nucleus.   
  
As a result of the analysis of the surface brightness profiles, we stress that except FS90-136,   
the remaining eight galaxies listed in Table\,2  will be considered as dE,N galaxies in the rest   
of this paper.   
  
\begin{table*}   
\begin{minipage}{170mm}   
\begin{center}   
\caption{Photometric/structural properties of a sample of Antlia dwarf elliptical galaxies.}    
\label{table2}   
\begin{tabular}{@{}lr@{}c@{}lr@{}c@{}lr@{}c@{}lr@{}c@{}lr@{}c@{}lr@{}c@{}lc@{}c@{}c@{}c@{}}   
\hline   
\\   
\multicolumn{1}{c}{ID}&\multicolumn{18}{c}{S\'ersic profile parameters}&  
\multicolumn{1}{c}{$m_{V,e}$}&\multicolumn{1}{c}{$m_{I,e}$}&\multicolumn{1}{c}{$m_{V,n}$}&\multicolumn{1}{c}{$m_{I,n}$}\\   
\multicolumn{1}{c}{}&\multicolumn{3}{c}{$\mu_{0,V}$}&\multicolumn{3}{c}{$r_{0,V}$}&\multicolumn{3}{c}{$N_V$}&  
\multicolumn{3}{c}{$\mu_{0,I}$}&\multicolumn{3}{c}{$r_{0,I}$}&\multicolumn{3}{c}{$N_I$}&\multicolumn{1}{c}{}&  
\multicolumn{1}{c}{}&\multicolumn{1}{c}{}&\multicolumn{1}{c}{}\\   
\multicolumn{1}{c}{}&\multicolumn{3}{c}{\tiny [${\rm mag arcsec^{-2}}$]}&\multicolumn{3}{c}{\tiny [${\rm arcsec}$]}&  
\multicolumn{3}{c}{}&\multicolumn{3}{c}{\tiny [${\rm mag arcsec^{-2}}$]}&\multicolumn{3}{c}{\tiny [${\rm arcsec}$]}&  
\multicolumn{3}{c}{}&\multicolumn{1}{c}{\tiny [${\rm mag}$]}&\multicolumn{1}{c}{\tiny [${\rm mag}$]}&
\multicolumn{1}{c}{\tiny [${\rm mag}$]}&\multicolumn{1}{c}{\tiny [${\rm mag}$]}\\   
   
\hline   
\\  
{\bf FS90-87}&$21.30$&$\pm$&$0.04$&$5.40$&$\pm$&$0.17$&$0.89$&$\pm$&$0.03$&$20.35$&$\pm$&$0.03$&$5.64$&  
$\pm$&$0.16$&$0.89$&$\pm$&$0.02$&$15.38$&$14.34$&$20.78$&$19.82$\\  
{\bf FS90-103}&$22.82$&$\pm$&$0.09$&$1.28$&$\pm$&$0.09$&$0.87$&$\pm$&$0.03$&$21.84$&$\pm$&$0.12$&$1.09$&  
$\pm$&$0.13$&$0.82$&$\pm$&$0.05$&$19.98$&$19.16$&$25.32$&$24.46$\\  
{\bf FS90-136}&$20.33$&$\pm$&$0.04$&$1.75$&$\pm$&$0.09$&$0.63$&$\pm$&$0.02$&$19.34$&$\pm$&$0.05$&$1.79$&  
$\pm$&$0.12$&$0.62$&$\pm$&$0.02$&$15.68$&$14.57$&$--$&$--$\\  
{\bf FS90-159}&$20.54$&$\pm$&$0.04$&$2.66$&$\pm$&$0.09$&$0.95$&$\pm$&$0.02$&$19.62$&$\pm$&$0.03$&$2.71$&  
$\pm$&$0.03$&$0.95$&$\pm$&$0.03$&$16.31$&$15.33$&$21.07$&$19.98$\\  
{\bf FS90-162}&$21.84$&$\pm$&$0.04$&$3.39$&$\pm$&$0.12$&$0.87$&$\pm$&$0.02$&$20.89$&$\pm$&$0.06$&$3.20$&  
$\pm$&$0.02$&$0.81$&$\pm$&$0.03$&$16.88$&$15.85$&$22.56$&$21.71$\\  
{\bf FS90-176}&$21.56$&$\pm$&$0.06$&$3.29$&$\pm$&$0.18$&$0.89$&$\pm$&$0.02$&$20.64$&$\pm$&$0.06$&$3.15$&  
$\pm$&$0.16$&$0.86$&$\pm$&$0.02$&$16.72$&$15.81$&$22.38$&$21.54$\\  
{\bf FS90-177}& & & & & & & & & & & & & & & & & & & & &$20.00$&$18.86$\\  
Inner comp.&$20.53$&$\pm$&$0.06$&$3.80$&$\pm$&$0.03$&$1.29$&$\pm$&$0.05$&$19.67$&$\pm$&$0.06$&$3.83$&  
$\pm$&$0.03$&$1.31$&$\pm$&$0.06$&$16.04$&$15.18$& &\\  
Outer comp.&$20.88$&$\pm$&$0.51$&$12.96$&$\pm$&$2.95$&$1.42$&$\pm$&$0.21$&$21.59$&$\pm$&$0.30$&$12.26$&  
$\pm$&$2.11$&$1.44$&$\pm$&$0.22$&$15.83$&$14.68$& &\\  
{\bf FS90-186}&$23.20$&$\pm$&$0.05$&$4.83$&$\pm$&$0.19$&$1.29$&$\pm$&$0.06$&$22.32$&$\pm$&$0.06$&$5.11$&  
$\pm$&$0.24$&$1.31$&$\pm$&$0.07$&$18.19$&$17.21$&$23.26$&$22.31$\\  
{\bf FS90-196}&$21.53$&$\pm$&$0.03$&$3.68$&$\pm$&$0.12$&$0.90$&$\pm$&$0.03$&$20.57$&$\pm$&$0.03$&$3.87$&  
$\pm$&$0.11$&$0.89$&$\pm$&$0.02$&$16.47$&$15.38$&$22.03$&$21.0$\\  

\hline   
\end{tabular}  
\end{center}    
NOTES: Columns $m_{V,e}$ and $m_{I,e}$ indicate the total integrated apparent magnitude of the external 
component in the V and I filter, respectively. In the case of FS90-177, the integrated magnitude of each 
one of the external components is shown separately. Columns $m_{V,n}$ and $m_{I,n}$ indicate the apparent 
magnitude of the nuclei in the V and I filter, respectively. At the adopted Antlia distance, $1\,{\rm arcsec}$
corresponds to $\approx 170\,{\rm pc}$.  
\end{minipage}   
\end{table*}

\begin{figure}   
\includegraphics[width=80mm]{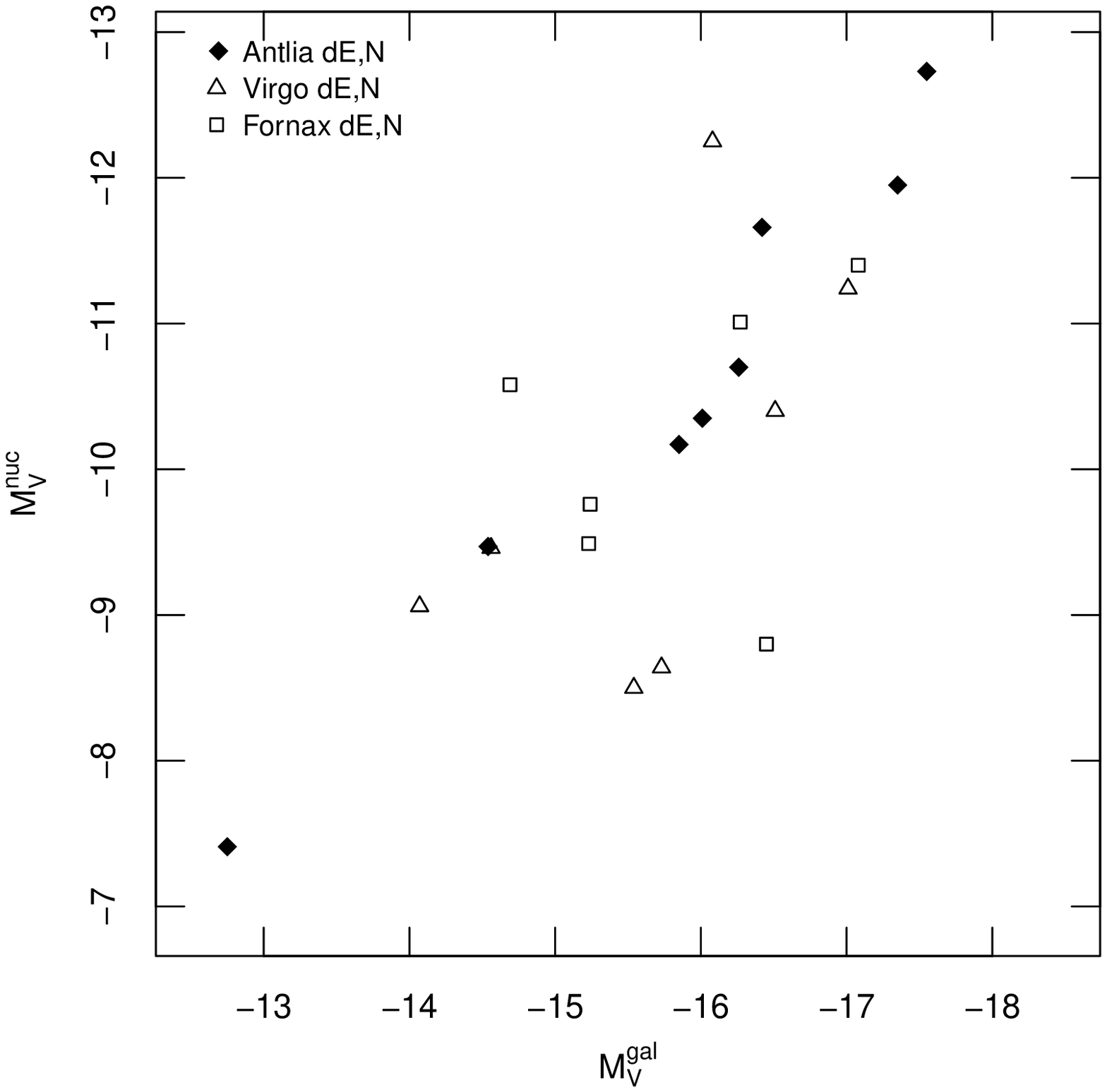}\\
\\   
\includegraphics[width=80mm]{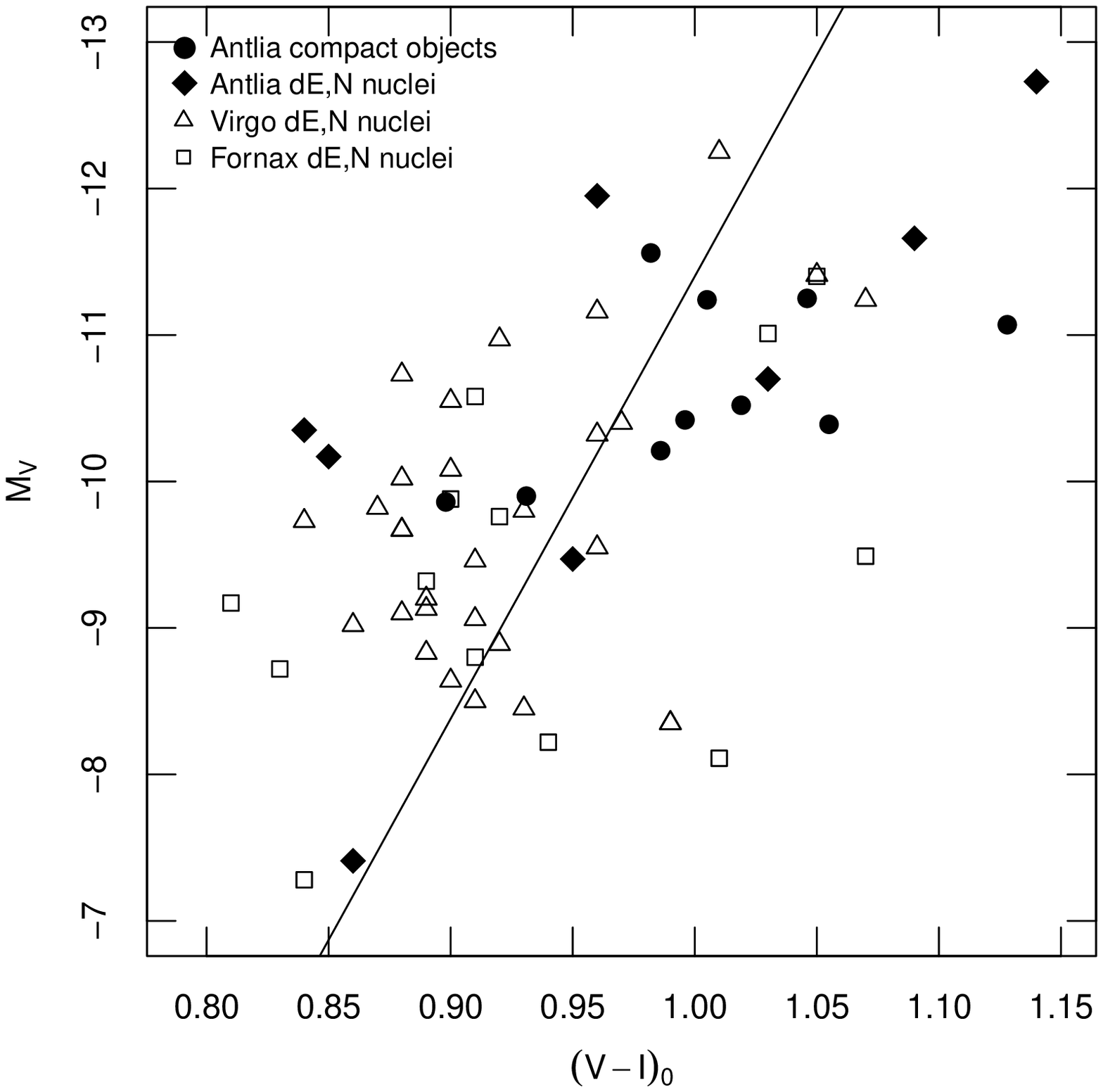}   
\caption{{\bf Upper panel}: Comparison of the total $M_V$ of dE,N galaxies in our Antlia   
sample with the $M_V$ of their respective nuclei (filled diamonds), including Virgo and   
Fornax dE,N galaxies (open triangles and squares, respectively, both from \citealt{lot04}).  
{\bf Lower panel}: Colour-magnitude $V,I$ diagram for Antlia compact objects (filled circles)   
and dE,N nuclei (filled diamonds). Open symbols represent dE,N nuclei from Virgo and Fornax   
(both from \citealt{lot04}). The solid line indicates the color-magnitude relation from 
\citet{smi12}, obtained for Antlia elliptical galaxies.}   
\label{den}   
\end{figure}   
  
The upper panel of Fig.\,\ref{den} shows the $M_V$ magnitude of the eight dE,N nuclei  
measured with our surface photometry versus the $M_V$ those of the host    
galaxies. For comparison, we include in this plot dE,N galaxies from the Virgo and Fornax   
clusters studied by \citet{lot04} whose $V$ magnitudes were available from \citet{sti01}.  
The galaxies in the three clusters seem to follow a similar trend, with brighter   
galaxies having brighter nuclei. Similar correlations can be detected in Fig.\,7 of \citet{lot04},   
where a larger sample of Virgo and Fornax nuclei are plotted versus $M_B$, as well as   
in \citet{cot06} where Virgo galaxies are studied within the ACS Virgo Cluster Survey.   
\citet{fer06} found for early-type Virgo galaxies a correlation between the masses of
nuclei and both host galaxy luminosity and virial mass.
 If a constant mass-to-luminosity ratio for   
dE,N nuclei is assumed, the correlation depicted in our plot would also point to   
a correlation between nuclei mass and galaxy luminosity.   
In addition, \citet{cot06} found that the luminosity ratio between nucleus and host galaxy
 does not depend on the galaxy luminosity, though a significant scatter is present in their 
relation. A similar conclusion can be drawn for our Antlia sample, if total luminosities are 
derived from the apparent magnitudes given in Table\,2.   
  
In the lower panel of Fig.\,\ref{den} we present the $V,I$ CMD of 10 ACOs 
listed in Table\,1 (those with $V,I$ data), 8 nuclei of our dE,N sample (Table\,2) as well   
as those from Virgo and Fornax taken from \citet{lot04}. The solid line represents the
CMR relation for Antlia confirmed elliptical galaxies from \citet{smi12}. In order to transform 
the CMR from Washington system to $V,I$, it was considered that $V-R \approx 0.6$ \citep{fuk95} and 
$(C-T_1)_0 = -0.65 + 2.04 \times (V-I)_0$. This last equation was derived from the relation given
by \citet{for01} for GCs, assuming that it can be applied to elliptical galaxies as well.
All compact objects and dE,N nuclei seem to occupy the same region in the CMD, presenting 
similar colours for similar luminosities. Even though the dispersion in the plot is large, 
they follow a common correlation where brighter objects tend to have redder colours. The 
CMRs defined by early-type cluster galaxies \citep[e.g.][and references therein]{smi12} present 
the same behaviour. Moreover, considering the uncertainties in the photometry, dE,N nuclei, 
dE,N nuclei and UCDs seem to follow the same trend as the \citet{smi12} CMR. For early-type galaxies, the metallicity 
is supposedly the parameter driving the relation, as more massive galaxies should be able to 
retain more metals during their evolution. However, \citet{pau11} have shown that the nuclei of 
early-type dwarfs in Virgo cover a large range in age, so it cannot be assumed that their colours 
depend just on metallicity. Taking into account both plots displayed in Fig.\,\ref{den}, it can 
also be suggested that brighter dE,N galaxies have brighter and redder nuclei, in agreement with 
the findings of \citet{cot06}.

\subsection{Effective radius and size-luminosity relation}
HST data permit to measure effective radii of extragalactic GCs or UCDs 
\citep[e.g.][]{mie07,mie08,evs08,mad10} out to the Antlia distance.
The knowledge of effective radii will result in a cleaner distinction of
GCs and UCDs. Moreover, new candidates may be identified, which could not be resolved
by ground-based data.
  
At the adopted Antlia distance, the ACS pixel size of 0.055\,arcsec represents   
$\sim\,9.3$\,pc, which is of the same order as the effective radius of small   
UCDs \citep{mie08,mis11}.  
  
\begin{figure*}   
\includegraphics[width=40mm]{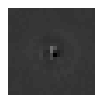}   
\includegraphics[width=40mm]{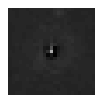}   
\includegraphics[width=40mm]{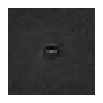}   
\includegraphics[width=40mm]{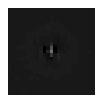}   
\includegraphics[width=40mm]{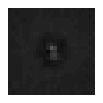}   
\includegraphics[width=40mm]{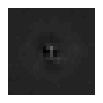}   
\includegraphics[width=40mm]{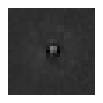}   
\includegraphics[width=40mm]{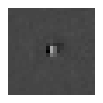}   
\caption{Residual maps from ISHAPE for the eight confirmed Antlia compact objects   
in the ACS field. From left to right, {\bf upper row:} ACOS\,2 to ACOS\,5.   
{\bf lower row:} ACOS\,6 to ACOS\,9  
}   
\label{ish}   
\end{figure*}   
   
In order to estimate the size of the confirmed Antlia compact objects located in the ACS field,   
ISHAPE \citep{lar99} was used for fitting their light profiles. Through this code, intrinsic shape   
parameters can be derived  for slightly extended sources, whose size   
is comparable to the FWHM of the Point Spread Function (PSF). It models    
the source as the convolution of an analytic profile and the PSF (obtained as   
explained in Section\,2.1). For each source, the output given by ISHAPE    
includes the FWHM of the object, the ratio of the minor over major axis, the    
position angle and the reduced $\chi^2$ parameter.   
   
From the list of analytic profiles offered by ISHAPE, we chose the King profile which accurately 
fits GC light profiles \citep{kin62,kin66}. Different concentration parameters ($c$, defined
 as the ratio of the tidal over the core radius) were applied, finding that for most of the 
compact objects the best $\chi^2$ estimation was obtained for $c=30$, in agreement with previous 
determinations \citep[e.g.][and references therein]{mad10}. It was also attempted to use $c$ as 
a free parameter, but most objects are only marginally resolved so that satisfactory fits could 
not be obtained.
   
ISHAPE was run on eight out of the eleven confirmed compact objects    
close to NGC\,3258, that are located in the ACS field, i.e. ACO\,2 to ACO\,9.    
Additionally, two foreground stars confirmed by radial velocities as well as another    
fourteen photometric UCD candidates were fitted. In this way, we could test    
the goodness of the ISHAPE output for these observations and obtain, in the    
case of objects without spectroscopic data, a refined list of candidates.    
   
\begin{figure}   
\includegraphics[width=84mm]{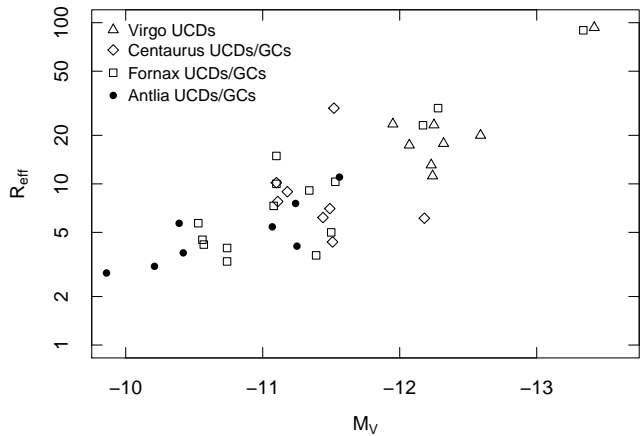}  
\caption{$R_{\rm eff}$ (logarithmic scale) versus $M_V$ for Antlia compact objects located    
close to NGC\,3258 in the ACS field, and UCDs/bright GCs from    
Fornax \citep{mie08}, Virgo \citep{evs08} and Centaurus \citep{mie07}.}   
\label{reff1}   
\end{figure}   
  
The ISHAPE estimation of the FWHM for the two foreground stars was $\sim 0.01$    
in pixel units, i.e. one tenth of the smallest FWHM obtained for the confirmed Antlia    
members. Considering that the    
FWHM of all the compact objects measured with ISHAPE in this investigation is    
less than one pixel, the eccentricities will not be taken into account. We should   
point out that there is no evidence for large projected ellipticities of GCs or 
UCDs (e.g. \citealt{har09} and references therein; \citealt{chi11}).   
The residual maps for Antlia confirmed members (Fig.\,\ref{ish}), given by ISHAPE show
clean subtractions.
  
Fig.\,\ref{reff1} shows the $R_{\rm eff}$ in logarithmic scale versus $M_V$, for the eight    
confirmed Antlia compact objects close to NGC\,3258 and located in the ACS field, together with   
similar objects from Fornax \citep{mie08}, Virgo \citep{evs08}, and Centaurus \citep{mie07}.    
 The brightest UCDs discovered in Virgo and Fornax, (VUCD\,7 and UCD\,3, according to the 
identification given by \citealt{evs08}) are also shown. Both of them, with   
$R_{eff}\,\sim\,100$\,pc, can easily be distinguished from the rest of the objects classified as   
UCDs/bright GCs, suggesting that their origin could in fact be different \citep{evs08}.    
The $R_{\rm eff}$ obtained for our sample of compact objects are in    
good agreement with those UCDs/bright GCs of similar absolute magnitudes studied in    
other clusters.

 The existence of a correlation between $log(R_{\rm eff})$ and luminosity   
for the brighter subsample has been discussed in the literature \citep{mie06,evs08,chi11,  
mad11,bro11}. Most authors support the existence of a size-luminosity relation for   
objects with $M_V < -11$ \citep[e.g.][]{mie06,mis11}) while no correlation is present for   
fainter objects (mainly GCs according to Mieske et al.). \citet{bro11} studied a sample 
of 34 confirmed UCDs around M\,87 with half-light radii of at least 10\,pc.
 They added a compilation from the literature of several different   
stellar systems, including objects of smaller sizes, and show a plot of half-light radii   
versus luminosity, where a break between UCDs and GCs (in their fig.\,8, lower panel) can   
be seen at a slightly different magnitude: $M_V \sim -10$, though the general   
dispersion makes it difficult a clear determination. In fact, Brodie et al. argue   
that UCDs do `not' show a clear size-luminosity relation, but different interpretations   
arise provided that the size and luminosity limits between GCs and UCDs are not   
widely agreed.   
 
In order to analyse the mean sizes in different magnitude ranges, we work with the joint 
sample of GCs/UCDs in Antlia, Centaurus, Fornax, and Virgo.
Dividing this whole sample into three magnitude ranges, the mean values of $R_{\rm eff}$   
for objects with $-13 < M_V < -12$, $-12 < M_V < -11$, and $-11 < M_V < -10$, are   
$17.9\pm2.4$\,pc, $9.8\pm1.5$\,pc, and $4.3\pm0.4$\,pc, respectively. That is, a trend is    
present with brighter objects having larger sizes.    
Considering a sample of 84 Galactic GCs with $-10 < M_V < -7$ \citep[][2010 Edition]{har96},    
a mean $R_{\rm eff}$ of $3.7\pm0.3$ is obtained. This is consistent, within the errors,     
with the result for our faintest range, showing that the transition in size of    
classic GCs towards brighter systems may be within or close to that magnitude bin.  
  
\begin{figure}   
\includegraphics[width=84mm]{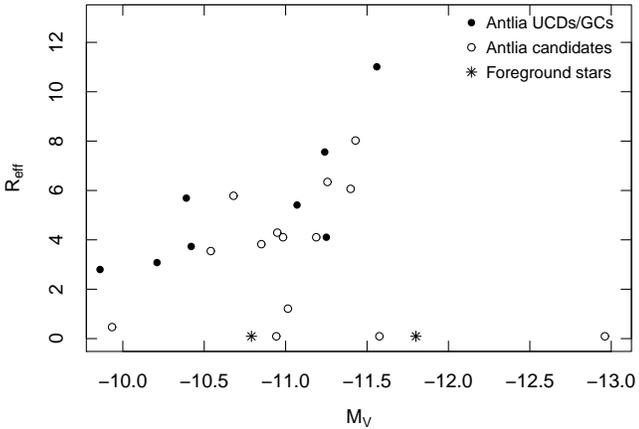}   
\caption{$R_{\rm eff}$ versus $M_V$ for compact objects close to NGC\,3258 that are Antlia    
confirmed members and candidates, plus two confirmed foreground stars, all of them located    
in the ACS field.}   
\label{reff2}   
\end{figure}   
   
As our Antlia sample reaches small $R_{\rm eff}$-values, it might serve to sharpen the size-luminosity 
relation. For this, we show in Fig.\,\ref{reff2} $R_{\rm eff}$ in a linear scale versus $M_V$ for the 
eight confirmed compact objects, two foreground stars, and the fourteen UCD/bright GC candidates 
located in the ACS field. The segregation between the confirmed compact objects and the foreground    
stars is very clear at $R_{\rm eff} \sim 1-2$\,pc. This gap allows us to clean the list of UCD/bright 
CG candidates from unresolved objects, i.e. probably foreground stars.    
Only nine out of the fourteen original candidates present effective radii in the same    
range as the confirmed compact objects do. The rest of them, according to the $R_{\rm eff}$    
given by ISHAPE, have low probabilities of being UCDs belonging to the Antlia    
cluster.
  
Now, we go back to the SExtractor's CLASS$\_$STAR and FWHM parameters that proved to be   
useful to select UCD/bright GC candidates. These nine marginally resolved UCD candidates   
selected according to the ISHAPE results, have been added as filled squares to the plots   
depicted in Fig.\,\ref{sex}.   
Moreover, framed squares and circles denote the confirmed members    
and candidates with $R_{\rm eff}>5$\,pc, respectively. In the left panel, the confirmed   
member with the largest $R_{\rm eff}$ has the lowest value of the SExtractor   
CLASS$\_$STAR=0.74, suggesting that it cannot be considered as a genuine point source.
 Adding these marginally resolved UCD candidates, the parameters CLASS$\_$STAR and 
FWHM show an acceptable capability to discriminate them from point sources. As 
expected, all compact objects with $R_{\rm eff}>5$\,pc present higher FWHM in the right 
panel.    
   
\begin{figure}   
\includegraphics[width=84mm]{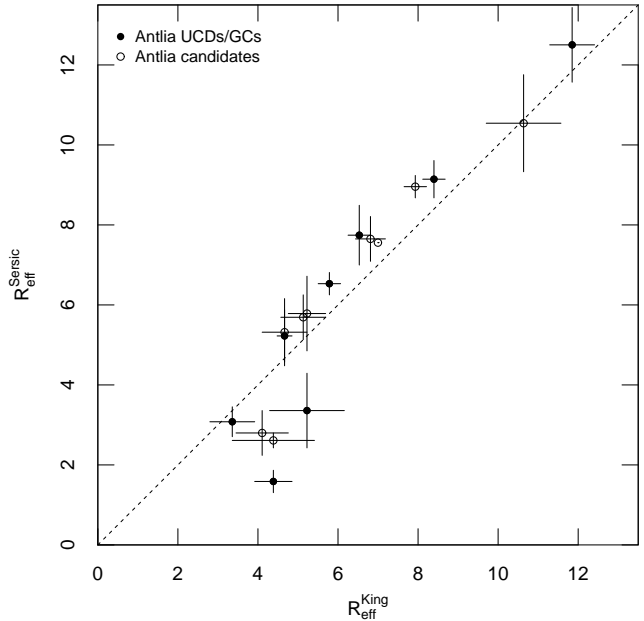}   
\caption{Comparison between the $R_{\rm eff}$ obtained with ISHAPE for confirmed Antlia    
members and marginally resolved candidates, applying a King profile (concentration    
parameter $c=30$) and a S\'ersic profile (index $n=2$).}   
\label{rvsr}   
\end{figure}

\citet{chi11} confirmed a sample of 27 UCDs as Coma cluster members, deriving their    
structural parameters with GALFIT \citep{peng02} and ISHAPE, employing S\'ersic profiles
and King profiles, respectively. They found a good agreement between the $R_{\rm eff}$ 
derived from both methods. Only the results for the largest object presents a clear 
discrepancy, but they attributed it to the presence of a halo component, that could 
affect the S\'ersic fits more than the King ones.
Therefore,  S\'ersic ($n=2$) fits were performed with ISHAPE to the Antlia compact objects 
located in the ACS field. The $R_{\rm eff}$ obtained with both analytic functions are 
depicted in Fig.\,\ref{rvsr}, showing a good consistency between the results within the 
errors. This suggests that none of these compact objects presents an extended, diffuse 
component and that the calculated $R_{\rm eff}$ are reliable and not model--dependent.     
  
If Fig.\,\ref{reff1} and Fig.\,\ref{reff2} are compared, it can be noticed that    
plotting the $R_{\rm eff}$ in a linear, not logarithmic scale, a break about   
$M_V \sim -10.5 - -11\,{\rm mag}$  could be seen, while for the fainter    
sources the $R_{\rm eff}$ seems to be independent of the visual absolute magnitude.      
As said above, a similar effect has been found by \citet{mie06} at $M_V \sim -11\,{\rm mag}$   
for the compact objects in Fornax. Mieske et al. also found a break in the 
metallicity distribution at this luminosity and argue that it can be 
considered as the limit between UCDs and GCs. \citet{chi11} studied the   
size-magnitude relation adding to the UCD Coma cluster members, a large amount of   
compact objects taken from the literature. They found that GCs in the range   
$-10<M_V<-8\,{\rm mag}$ have nearly constant half-light radii, independently of their   
luminosity, but the compact objects brighter than $M_V=-10\,{\rm mag}$ (i.e. in the UCD luminosity   
regime) display a trend of increasing size with increasing luminosity.   
A similar break about $M_V=-10\,{\rm mag}$ can be seen in the half-light radius versus luminosity   
plot presented by \cite{mis11} for Hydra\,I compact objects in addition to other star   
clusters and UCDs.   
  
This break in the size-luminosity plane, attributed to the limit between GCs and UCDs,   
does not appear to be well defined as in the literature it spans integrated magnitudes   
$M_V=-10 - -11\,{\rm mag}$. However, it should be taken into account that the existence of such   
an `abrupt' change in the slope of   
this relation is quite unlikely, as the evidence points to a magnitude range where both   
UCDs and GCs coexist, and also the distance and photometric errors are affecting directly   
the absolute magnitude determination.   
  
\section{Properties of the UCD sample}   
   
After studying the sizes of the Antlia compact objects lying within the ACS field   
and measuring radial velocities that confirm several cluster members,   
it is possible to define a new sample with the most reliable selection of   
UCD Antlia members and candidates, which we will refer to in the rest of the paper.   
Among them, we have chosen those that are   
brighter than $M_V \sim -10.5\,{\rm mag}$, that is the luminosity limit we adopt for this `final   
UCD sample'. In this way, we gather six ACO objects from Table\,1 (i.e. confirmed by radial   
velocities) plus the nine candidates located in the ACS field that are marginally resolved   
sources according to their light profiles, all of them with $M_V \leq -10.5\,{\rm mag}$. This 
leaves us with 15 UCDs that are Antlia members or candidates around NGC\,3258.   
  
\subsection{Projected spatial distribution through Principal Component Analysis}      
  
\begin{figure}   
\includegraphics[width=84mm]{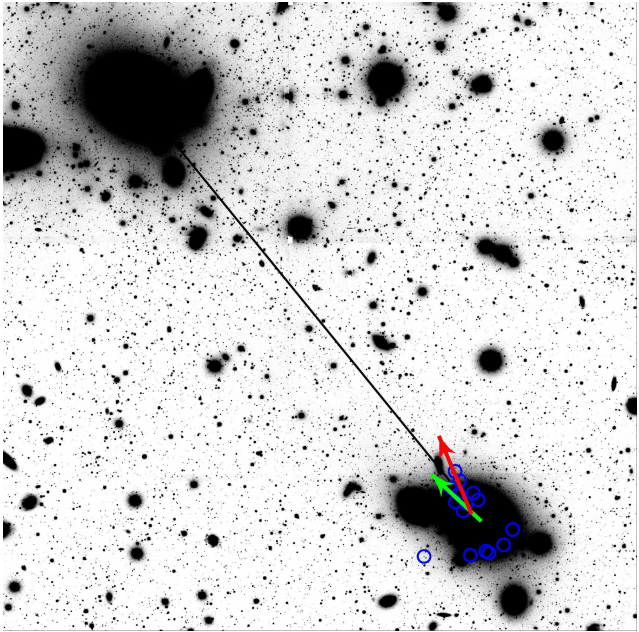}   
\caption{Projected spatial distribution of the objects in the `final UCD sample' (blue open 
circles). The red arrow indicates the principal component direction obtained from the PCA for 
the UCD sample, while the green one indicates the principal component direction obtained for a 
GC candidates sample ($M_V>-10.5$ and $0.75< V-I <1.4$, galactocentric radius 
$r \approx 0.5-2\,{\rm arcmin}$). The 
black solid line that joins NGC\,3258 and NGC\,3268 centers is added for comparative purposes. 
North is up, East to the left.}   
\label{pca}   
\end{figure}

In order to compare the projected spatial distribution of the UCD sample with   
that of the GC candidates surrounding NGC\,3258, we apply Principal Component    
Analysis (PCA).  
   
PCA is a very popular tool in modern data analysis, used for a diverse field of   
problems because it has the ability of reducing a complex    
data set to a lower dimension, revealing the structures that sometimes underlie it. PCA    
uses an orthogonal transformation to  convert a set of observations of possibly correlated    
variables into a set of values of uncorrelated variables, less than or equal to the    
number of original variables, called principal components. As a consequence of the    
transformation, the greatest variance by any projection of the data will lie on    
the first coordinate of the new coordinated system (first eigenvector), the second    
greatest variance on the second one, and so on. For more details on PCA, we refer to    
\citet{jol02}.   
   
A PCA has been run on the spatially projected coordinates of the UCD sample, using 
the task PRINCOMP from the software R Project for Statistical Computing \citep{rde11}. 

 The position angle (PA) of the first eigenvector (PA\,$\sim 23\degr$, 
red arrow in Fig.\,\ref{pca}), is slightly different than that of the corresponding 
eigenvector obtained for the GC candidates ($M_V>-10.5$ and $0.75< V-I <1.4$) within the 
radial range 0.5--2\,arcmin from NGC\,3258 (PA\,$\sim 39 \degr$, green arrow in 
Fig.\,\ref{pca}). 
This latter PA is in agreement with those obtained by \citet{dir03b}    
and \citet{bas08} studying the azimuthal distribution of the NGC\,3258 GCS,    
i.e. $38\degr \pm 6\degr$ and $32\degr \pm 5\degr$, respectively.    
Taking into account the small size of the UCD sample, it can be considered   
that the results of PCA of both samples, UCDs and GCs, show similar   
preferential directions for their projected spatial distributions. In addition, they also 
agree with the PA with origin in NGC\,3258 pointing to the direction of NGC\,3268, of 
$39 \degr$ \citep[][, black solid line in Fig\,\ref{pca}]{bas08}.   
  
Thus, the GCS of NGC\,3258 is elongated in the direction towards NGC\,3268, and a similar   
effect is present in the X-ray emission around NGC\,3258, that has an extension in the   
same direction. The results of the PCA suggest that UCDs also present a projected distribution   
with a similar orientation. All this evidence lends support to the idea that tidal   
forces, between both giant ellipticals, may be playing an important role in modelling   
the spatial distribution of the hot intra-cluster gas and stellar systems at the Antlia   
cluster core. \citet{smi12} have shown that each giant is surrounded by a retinue of   
normal and dwarf early-type galaxies, and this scenario may correspond to an ongoing merger   
between two groups in which all, i.e. galaxies, UCDs, GCs, and hot gas, seem to participate.   
In fact, preliminar results of the X-ray study performed by \citet{haw11} lend support to   
the idea that Antlia is a galaxy cluster at a phase of an intermediate merger.   
  
\subsection{Colour distribution}   
  
\begin{figure}   
\includegraphics[width=84mm]{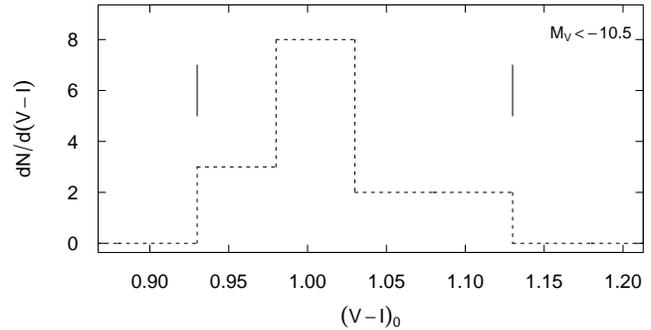}   
\caption{Colour distribution for UCDs close to NGC\,3258, corresponding to Antlia    
confirmed members plus candidates located in the ACS field, with $M_V <-10.5$   
(final UCD sample). Solid lines indicate the colours of the peaks for the NGC\,3258 
blue and red GC populations from \citet{bas08}.}   
\label{dcol}   
\end{figure}   
  
Fig.\,\ref{dcol} shows the colour distribution for the UCD sample. Clearly, there   
is a high fraction of UCDs with $(V-I)_0 \sim 1.0$, with relatively low dispersion.   
For the inner region of the NGC\,3258 GCS, \citet{bas08} fitted two Gaussians   
to the GC colour distribution and obtained peaks at $(V-I)_0=0.93\pm0.01$ for   
the blue GCs, and at $(V-I)_0=1.13\pm0.01$ for the red ones, being the colour limit between   
them $(V-I)_0 = 1.05$. This would place the colour distribution of our UCD sample closer to   
the colour range of the blue GCs. A KMM test was applied, using the algorithm described   
by \citet{ash94}. Considering as the null hypothesis that the UCD sample colour distribution   
is better represented by the sum of two homoscedastic (i.e. with similar dispersion) Gaussians   
than by a single Gaussian, the result indicates that this hypothesis cannot be rejected, with   
a $90\%$ of confidence. However, the results of this KMM test point out that the existence of   
the second (redder) Gaussian is based only on the two redder members of the sample. These are the   
only UCDs for which the algorithm estimates a probability of belonging to the red subpopulation   
significantly different from zero. Then, we decided to rerun the KMM test excluding these two red   
objects from the UCD sample, and then the null hypothesis was rejected. According to these results, the   
colour distribution of UCDs in the vicinity of NGC\,3258 is mainly described by a single Gaussian,   
with the exception of the two redder members.   
Their colours are $(V-I)_0 \leq 1.05$, i.e. within the colour range of blue GCs and in agreement with   
the results obtained at Section\,3.1 for the preliminar sample, where fainter objects (probable GCs)   
were also included.   
  
Although our Antlia sample is rather small, we note that the majority of the UCDs, i.e. 13 out of the 
15 discovered ones, have colours corresponding to the blue GC range, as well as 6 out of the 8 dE,N 
nuclei studied in the same cluster. In other clusters, UCDs cover the whole GC range as shown for 
instance by \citet{evs08} for Virgo and Fornax, and by \citet{mad10} and \citet{chi11}   
for Coma. On the other hand, UCDs in the Hydra cluster mostly appear as an extension of the red GC   
subpopulation towards brighter magnitudes \citep{weh07}. Such colour differences can be understood   
as another evidence that UCDs have multiple origins, even in similar environments, and there is   
no single theory that could explain all of them \citep{nor11}. We will come back to this issue in   
Section\,4.4.

\subsection{Antlia colour-magnitude diagram}  
  
\begin{figure}   
\includegraphics[width=84mm]{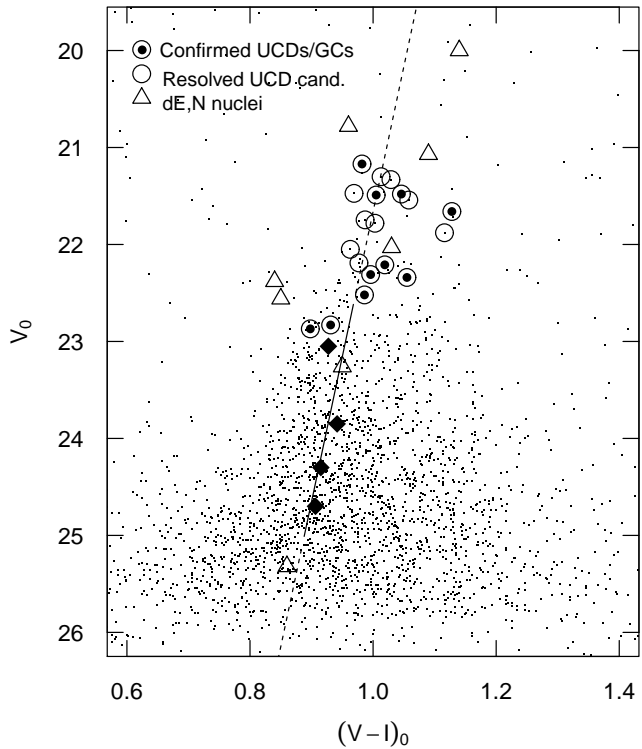}  
\caption{Antlia $(V,I)$ CMD corresponding to the faint luminosity regime,    
where GCs, UCDs, and dE,N nuclei are plotted. Open circles indicate UCD candidates (i.e., without 
radial velocity measurements) marginally resolved in the ACS field, while framed filled circles
indicate the confirmed ones. Dots are GC candidates from \citet{bas08} and open triangles are dE,N 
nuclei from this work. Filled diamonds are the mean colours of blue GCs in four magnitude
bins. The solid line represents the fit to these points (''blue tilt''), while the dashed lines indicate its
extrapolation.}   
\label{cmdall}   
\end{figure}

Fig.\,\ref{cmdall} shows the Antlia $(V,I)$ CMD corresponding to the faint luminosity regime,    
where GCs, UCDs, and dE,N nuclei are plotted.  The GC photometric data were obtained   
from \citet{bas08} and the rest of the data from this paper.   
  
It has already been noticed by \citet{har06}, in an ACS--HST photometric study, that bright blue GCs   
in NGC\,3258 and other brightest cluster galaxies present a correlation in the CMD, in the sense   
that they get redder with increasing luminosity, i.e. the currently called `blue tilt'   
\citep[e.g.][]{str06,bro06}.    
In a later $(V,I)$ study of the NGC\,3258 GCS by \citet{bas08}, the blue tilt is still perceptible   
in the CMD (their fig.\,6) though it is not specifically examined, and it can also be detected   
at Fig.\,\ref{cmdall} as the high luminosity limit of the GCs. In order to quantify this effect,
NGC\,3258 GCs with $22.5<V_0<25$ where separated in four equally populated magnitude bins. For each range, we
obtained the background-corrected colour distribution and fitted the sum of two Gaussians to it. The
solid line in Fig.\,\ref{cmdall} is the result of fitting the mean values of the blue GCs in each bin,
while the dashed lines indicate its extrapolation. This corresponds to a `blue tilt' slope of 
$d(V-I)_0/dV_0 \approx -0.03$. In this latter figure, the framed filled   
circles show all the Antlia members confirmed with radial velocities (Table\,1), including both UCDs   
and GCs. The empty circles correspond to the UCD candidates located in the ACS field that are marginally   
resolved. The majority have colours in the range of blue GCs (bluer than $(V-I)_0 \sim 1.05$) except   
two objects with $(V-I)_0 > 1.10$ that are located on the `red' side. Those identified with large circles   
on the `blue' side follow the same trend as the blue tilt does; in fact, some of the faintest ones   
probably belong to the GC population.   
  
One of the most accepted explanations of the GC blue tilt is that it corresponds to a mass-metallicity   
relation of metal-poor clusters, mainly driven by chemical self-enrichment \citep[e.g.][]{mie10,for10}.   
Moreover, the possibility of a pre-enrichment process, or a combination of both, has also been addressed   
with models by \citet{bai09}.  
  
The dE,N galaxies whose nuclei are depicted in Fig.\,\ref{cmdall} are Antlia members located on our   
four VLT fields; that is, not all of them are close to NGC\,3258 but within the cluster core. Most of 
the dE,N nuclei in the CMD share the same location, on the blue side, with UCDs and GCs of our sample,   
with the exception of one nucleus that is blue but much fainter (corresponding to the faintest dwarf
galaxy in our sample), and another one that is clearly on the red side and has the brightest magnitude 
of the sample. This sequence is also in agreement with that followed by Fornax and Virgo dE,N nuclei 
displayed at the lower panel of Fig.\,\ref{den}.    
  
\citet{cot06} compared Virgo dE,N nuclei with UCDs in the same cluster from \citet{has05},   
and they point out that there is agreement between both samples in terms of colour,   
luminosity and size. Also in Virgo, \citet{pau10} performed a comparison of the stellar population   
of a sample of UCDs and dE,N nuclei. They obtained different metallicities and stellar population ages   
for the UCDs and the general nuclei sample. However, if their dE,N sample is restricted to those   
located in the high-density regions of Virgo, where most of the UCDs are also located, the authors   
reported that no significant differences are found between the populations of both kind of objects,   
being all old and metal-poor. Taking this into account, they suggest that Virgo UCDs may in fact be   
dE,N remnants.   
  
For the Antlia cluster, we find a similar agreement regarding colours and luminosities between   
dE,N nuclei and mainly blue UCDs. All this 
evidence points to a close relation between them, as   
already mentioned by \citet{har06}, that suggests that blue UCDs may be the remnants of disrupted   
dE,N galaxies, i.e. stripped dwarf galaxy nucleus that may be the result of minor merger events.   
This idea has been sustained since the original models performed by \citet{bas94} and later on by   
\citet{bek01}, among others, until some of the most recent observational evidence given by e.g.   
\citet{bro11} and \citet{nor11}. However, the same origin does not seem to apply for red UCDs,   
that appear to be in a minority close to NGC\,3258 in Antlia, as we just find two confirmed ones.   
For instance, \citet{nor11} propose that red UCDs are just the most luminous GCs associated to   
the host galaxy, while \citet{bro11} add a third scenario and suggest that some red UCDs may be   
related to the remnants of more massive and metal-rich galaxies. We can not discard that part of
the UCDs in our sample may belong to the bright end of the NGC\,3258 GCS.

\section{Summary and conclusions}   
On the basis of images obtained with FORS1, MOSAIC, and the ACS--HST archive,   
as well as radial velocities measured on GMOS--GEMINI spectra, we have studied the first   
`compact objects' discovered in the Antlia cluster, that comprise UCD members, UCD    
candidates and bright GCs located in the surroundings of the giant elliptical NGC\,3258.   
Particularly, UCD members and candidates are selected among these compact objects, on the   
basis of luminosity, colour, and size. Our main results and conclusions are summarized   
in the following.  
  
\begin{itemize}  
  
\item The Antlia compact objects comprise eleven new members (Table\,1), named with the   
acronym `ACO', whose cluster membership is kinematically confirmed with our GMOS spectra.   
Except for one object that lacks $V,I$ photometry, they follow a CMR similar to that defined by Fornax   
and Virgo compact objects, but they do not reach luminosities as high as their counterparts in   
the other clusters (Fig.\,\ref{cmd}). They have $M_V > -11.6\,{\rm mag}$ and colours mostly in the   
range corresponding to blue GCs.
  
\item We have obtained surface brightness profiles of nine dE galaxies in the range 
$-18 < M_V < -13\,{\rm mag}$ from the FS90 catalog \citep{fer90}, that are Antlia confirmed   
members \citep{smi08a,smi12}. S\'ersic models have been fitted to them and the best fit parameters   
(central surface brightness, scale parameter, and S\'ersic index) as well as total $V$ and $I$   
magnitudes of the different components have been determined. Out of these nine galaxies, seven have   
two components (a nuclear component plus a halo), one has three components (a nucleus plus two   
components necessary to fit accurately the outer halo), while the other one has only one component.   
Thus, we can confirm their classification as one non-nucleated dE (FS90-136) and eight nucleated   
dE,N galaxies.   
  
\item We verify that absolute magnitudes of the Antlia dE,N stellar nuclei correlate with total   
absolute magnitudes of their host galaxies, so that brighter galaxies tend to have brighter nuclei,   
in a similar way as has been detected in Fornax and Virgo \citep[e.g.][]{lot04,cot06}. If the nuclei     
have a constant mass-to-luminosity ratio, then this relation would point to the existence of a nuclei   
mass versus host galaxy luminosity (and mass) correlation, as found by \citet{fer06} in the Virgo cluster.   
  
\item The nuclei of dE,N Antlia galaxies share the same locus in the CMD with nuclei of dE,N   
galaxies from Fornax and Virgo (Fig.\,\ref{den}). They all get redder with increasing luminosity,   
following a colour-luminosity relation that was first noticed by \citet{cot06} (see also \citealt{pau11}).   
With regard to colours, few nuclei in this sample from Antlia, Fornax, and Virgo,   
are redder than $(V-I)_0 = 1.05$, i.e. the adopted limit between blue and red GC.    
However, this may be a consequence of the selected sample being not bright enough,   
as brighter nuclei tend to be redder. \citet{pau11} suggest that the processes that govern   
the formation of nuclei might be quite different for bright or faint dE galaxies. According   
to Paudel et al., the nuclei of faint dEs have old and metal-poor populations,   
while those of bright dEs are younger and more metal-rich. 
The Antlia compact objects ($-11.6 > M_V > -9.5\,{\rm mag}$) are located at the same position in the CMD   
as the dE,N nuclei from Antlia, Fornax, and Virgo (Fig.\,\ref{den}).   
  
\item By means of the ISHAPE code, we have determined the sizes of eight Antlia compact objects   
located on ACS images. They have effective radii in the range $R_{\rm eff} = 3 - 11$\,pc.   
These objects together with nine new photometric candidates with effective radii within the same   
range constitute our `UCD sample' close to NGC\,3258, i.e. our most reliable selection.   
A $R_{\rm eff} \sim 2$\,pc is taken as the lower limit for Antlia UCDs, where a clear gap is   
present in the $R_{\rm eff}$ versus $M_V$ plot (Fig.\,\ref{reff2}). Two foreground stars, confirmed   
with radial velocities, are included in this plot and show that point sources are located below such   
limit. We remind that GCs are detected as point sources on the ACS images at the Antlia distance.   
  
\item The Antlia UCD sample as well as UCDs/bright GCs in Fornax, Virgo, and Centaurus show the same   
behaviour regarding the size-luminosity relation. For objects brighter than a limiting magnitude,   
the $R_{\rm eff}$ seems to increase with increasing luminosity, while for fainter magnitudes the   
$R_{\rm eff}$ remains almost constant. We find this limiting magnitude or `break' in the size-luminosity   
relation at $M_V \sim -10.5 - -11\,{\rm mag}$, though different authors set it at slightly different   
magnitudes \citep[e.g.][]{mie06,chi11,mis11}.   
  
\item The projected spatial distribution of the UCD sample has similar characteristics as those of the   
NGC\,3258 GCS and X-ray emission. They all present a PA in the direction to the other giant elliptical   
that dominates Antlia, NGC\,3268. These pieces of observational evidence point to an ongoing merger   
between two groups in Antlia. Future kinematic studies will help to settle the question.    
  
\item  Most UCDs in our sample have colours within the range defined by the blue GCs, and only two   
appear on the red side. The blue ones follow a CMR similar to the blue tilt defined by the brightest   
blue GCs \citep{str06,bro06}, getting redder with increasing luminosity. Moreover, six out the eight   
Antlia dE,N nuclei share the same locus on the CMD. As a consequence, we propose that some blue UCDs around 
NGC\,3258 may be the remnants of stripped dwarf galaxies captured during minor merger events. Regarding   
the red ones, they seem to be much less numerous than the blue ones in this location, and we speculate   
that they may be part of a UCD subpopulation whose origin/s is/are different from the blue ones   
\citep[e.g.][]{bro11,nor11}. We plan to extend our research on Antlia UCDs to the rest of the cluster   
in order to gather more evidence to deepen on these hypotheses.     
  
\end{itemize}

\section*{Acknowledgments}  
We thank the anonymous Referee for useful comments that improved the original version.
This work was funded with grants from Consejo Nacional de Investigaciones   
Cient\'{\i}ficas y T\'ecnicas de la Rep\'ublica Argentina, Agencia Nacional de   
Promoci\'on Cient\'{\i}fica y Tecnol\'ogica, and Universidad Nacional de La Plata  
(Argentina). TR is grateful for financial support from FONDECYT project Nr.\,1100620, 
and from the BASAL Centro de Astrof\'isica y Tecnolog\'ias Afines (CATA) PFB-06/2007.
ASC acknowledges finantial support from Agencia de Promoci\'on Cient\'ifica y 
Tecnol\'ogica of Argentina (BID AR PICT 2010-0410).

\label{lastpage}   
\end{document}